\documentclass[]{aastex631}

\usepackage{booktabs}
\usepackage{subfigure}
\usepackage{amsmath}
\usepackage{multirow}
\usepackage{array}
\usepackage{soul}
\usepackage[utf8]{inputenc}

\usepackage{gensymb}

\submitjournal{AJ}

\shorttitle{Estimation of Photometric Redshifts}
\shortauthors{Lee \& Shin}

\begin{document}

\title{Estimation of Photometric Redshifts. II. Identification of 
Out-of-Distribution Data with Neural Networks}

\author[0000-0003-2510-4132]{Joongoo Lee}
\affiliation{Korea Astronomy and Space Science Institute, 776,
Daedeokdae-ro, Yuseong-gu,                                                      
Daejeon 34055, Republic of Korea}
\affiliation{Department of Physics and Astronomy, 
Seoul National University, 1, Gwanak-ro, Gwanak-gu,
Seoul 08826, Republic of Korea}
\email{goolee5286@gmail.com}

\author[0000-0002-9934-3139]{Min-Su Shin}
\correspondingauthor{Min-Su Shin}
\affiliation{Korea Astronomy and Space Science Institute, 776,
Daedeokdae-ro, Yuseong-gu,                                                      
Daejeon 34055, Republic of Korea}
\email{msshin@kasi.re.kr}

\begin{abstract}
In this study, we propose a three-stage training 
approach of neural networks for both photometric redshift estimation of galaxies 
and detection of out-of-distribution (OOD) objects. 
Our approach comprises supervised and unsupervised 
learning, which enables using unlabeled (UL) data for OOD detection 
in training the networks. 
Employing the UL data, which is the dataset most similar to the real-world data, 
ensures a reliable usage of the trained model in practice. We 
quantitatively assess the model performance of photometric redshift 
estimation and OOD detection using in-distribution (ID) galaxies 
and labeled OOD (LOOD) samples such as stars and quasars. 
Our model successfully produces 
photometric redshifts matched with spectroscopic redshifts for the ID 
samples and identifies well the LOOD objects with more than 98\% accuracy. 
Although quantitative assessment with the UL samples is impracticable due to the lack of 
labels and spectroscopic redshifts, we also find that our model 
successfully estimates reasonable photometric redshifts for ID-like 
UL samples and filter OOD-like UL objects. 
The code for the model implementation is available at 
\href{https://github.com/GooLee0123/MBRNN\_OOD}{https://github.com/GooLee0123/MBRNN\_OOD}.
\end{abstract}

\section{Introduction \label{sec:intro}}

Future astronomical surveys, such as the Legacy Survey of Space and 
Time \citep{2019ApJ...873..111I}, Euclid \citep{10.1117/12.857123}, 
and Nancy Grace Roman Space Telescope \citep{2013arXiv1305.5422S}, 
are expected to explore billions of galaxies whose redshifts should be 
reliably estimated. The redshifts of galaxies 
are prerequisites in most extragalactic and cosmological 
studies 
\citep{10.1111/j.1365-2966.2005.09526.x,10.1111/j.1365-2966.2010.16765.x,2015MasterAJ}. 
Although spectroscopic measurements yield accurate redshifts of 
distant galaxies with negligible error \citep{2016MNRAS.460.1371B}, 
obtaining spectroscopic redshifts is excessively time-consuming. Moreover, the 
number of objects to which spectroscopy is applicable is limited because 
high-quality spectroscopic data are high-priced to acquire particularly for 
faint objects. The 
cost of spectroscopic redshifts leads to the use of photometric redshifts 
as alternatives. 
Photometric redshifts are 
less time-intensive and more widely applicable than 
spectroscopic ones, although their accuracy is 
generally worse than that of spectroscopic redshifts \citep{2019NatAs...3..212S}.

Template-based spectral energy distribution (SED) 
fitting \citep{2000AA...363..476B} and machine-learning inference 
are two representative approaches for estimating photometric redshifts 
\citep{2017MNRAS.465.1959C}. Although the template-based 
SED fitting provides extensive coverage of redshifts and 
photometric properties, it can be more computationally expensive than 
machine-learning inference due to the nature of brute-force search. Besides, the 
reliance on prior knowledge on SEDs may induce bias in 
results \citep{2011ApSS.331....1W,2015ApJ...801...20T}. Meanwhile, 
machine-learning approaches enable quick inference of 
photometric redshifts after training phases \citep{2019NatAs...3..212S} 
as proven in many past studies 
\citep{Carliles_2010,Gerdes_2010,2018AA...609A.111D,2018AA...611A..97P}. 
Despite this advantage of the machine-learning approaches, 
the reliability and performance of machine-learning 
models are guaranteed only when test data follow the 
distribution of training data in 
both input and target (i.e., redshift) spaces as 
in-distribution (ID) data \citep{DBLP:journals/corr/GoodfellowSS14,AmodeiAISafety}.

It is practically impossible to prepare a training set with 
complete coverage of the entire input and target (i.e., redshift) spaces 
extended by the expected real-world data \citep{Domingos12cacm,
YuilleL21}. The under-represented or unseen 
galaxies residing outside the property range of training samples for inference of 
photometric redshifts are drawn from out-of-distribution (OOD) with 
respect to given machine-learning models 
\citep{DBLP:conf/iclr/HendrycksG17,liang2017enhancing,lee2018training,
NEURIPS2018_abdeb6f5,NIPS2019_9611}. 
The OOD objects in the inference of photometric redshifts 
can be categorized into two types. One type 
corresponds to objects such as quasars (QSOs) and stars physically 
different from galaxies. We call them physically OOD objects. 
Although the distribution of 
the physically OOD samples generally deviates from that of ID samples 
in the input space, some may exist where the ID samples 
densely populate. For example, the colors of QSOs at some redshifts 
can be similar to those of galaxies at different redshifts. 
Hence, the physically OOD samples may be 
viewed as ID samples from the perspective of the trained model. 
The second type comprises galaxies unseen or under-represented 
in training data (hereinafter, under-represented galaxies). 
The photometric redshifts of under-represented galaxies cannot 
be successfully inferred by a model due to the nature of inductive 
reasoning in machine-learning.

In our previous study (Paper I, \citet{paperI}), we focused on the accurate 
photometric redshift estimation of ID samples that are galaxies well represented 
by training data, and the trained model successfully inferred 
photometric redshifts with ID test samples. 
However, we found that both non-galaxy objects and 
under-represented galaxies could have an overconfident estimation of photometric redshifts, 
although the estimation was incorrect. 
Confidence is a measure of 
how confident a neural network is about its inference, and it is defined 
as the maximum value of the probability output for a given sample. Silent 
failure of trained networks by estimating incorrect photometric redshifts 
with high confidence for the OOD objects proves that the typical confidence 
is an unreliable indicator for detecting OODs and may lead to 
wrong interpretation \citep{7298640}. 
Owing to these unreliable results and the vulnerability of trained 
networks, alternative approaches for OOD detection are required.

In this second paper of a series, we propose a multi-stage-training 
approach to 
handle dual tasks: photometric redshift estimation and OOD 
scoring/detection\footnote{We interchangeably 
use OOD scoring and detection. Specifically, we use the term 
{\it scoring} when emphasizing the sequential property of OOD score and 
use {\it detection} otherwise.}. The proposed approach 
employs unsupervised learning for OOD scoring/detection introduced in 
\citet{Yu_2019_ICCV} in addition to supervised learning for redshift estimation described 
in Paper I. Supervised learning requiring labeled samples (i.e., training samples of 
galaxies with spectroscopic redshifts) in training the model 
shows a remarkable performance on ID 
data \citep{10.1145/1143844.1143865} (see Paper I). 
We employ unsupervised learning, which 
trains models with a loss function computable without 
sample labels \citep{10.1145/502512.502570,1262178} for using unlabeled (UL) data, 
presumably containing both ID and OOD data. 
For inference, we use networks' outputs from different timelines of 
the training process to evaluate photometric redshifts and OOD scores. Since 
the networks trained as such can both detect OOD 
objects and estimate photometric redshifts, we expect 
a trained model to perform better with real-world data than a model 
without handling the OOD samples.

We also expect that our approach employing the UL training data for OOD object 
detection can achieve a more robust performance regarding the OOD data by 
handling the more diverse possibility of OOD properties with 
the UL data than the labeled data. Our trained networks may not be capable of 
detecting OOD objects populating the regions of the input spaces not covered 
by the UL data. However, the regions that UL data encompass in the input space 
can be easily extended because they obviate high costs in labeling training samples. 

The rest of the paper is organized as follows. In 
Section \ref{sec:data}, we introduce ID, labeled OOD (LOOD), and UL 
datasets used for the training and inference of networks. 
Section \ref{sec:method} describes the proposed multi-stage-training 
approach for the two tasks: photometric 
redshift estimation and OOD detection. Section \ref{sec:result} focuses on 
the performance analysis of the networks for both tasks using 
ID, LOOD, and UL test datasets. Finally, we summarize our contribution 
and discuss the future research direction in Section \ref{sec:discon}.

\section{Data \label{sec:data}}

\begin{table*}
\centering
\caption{Spectroscopic galaxy redshift samples. \label{tab:spec_z_samples}}
  \hspace*{-2.5cm}\begin{tabular}{c|c p{6cm} c}
\tableline\tableline
Dataset name & Number of objects & Selection conditions & Reference \\
\tableline\tableline
  SDSS DR15 & 1294042 & (CLASS $==$ GALAXY) and (ZWARNING $==$ 0 or 16) and (Z\_ERR $>=$ 0.0) & \citet{2019ApJS..240...23A} \\
  LAMOST DR5 & 116186 & (CLASS $==$ GALAXY) and (Z $>$ -9000) & \citet{2012RAA....12.1197C} \\
  6dFGS & 45036 & (QUALITY\_CODE $==$ 4) and (REDSHIFT $<=$ 1.0) & \citet{2009MNRAS.399..683J} \\
  PRIMUS & 11012 & (CLASS $==$ GALAXY) and (ZQUALITY $==$ 4) & \citet{2013ApJ...767..118C} \\
  2dFGRS & 7000 & (Q\_Z $>=$ 4) and (O\_Z\_EM $<$ 1) and (Z $<$ 1) & \citet{2001MNRAS.328.1039C} \\
  OzDES & 2159 & (TYPES $!=$ RadioGalaxy or AGN or QSO or Tertiary) and (FLAG $!=$ 3 and 6) and (Z $>$ 0.0001) & \citet{2017MNRAS.472..273C} \\
  VIPERS & 1680 & (4 $<=$ ZFLG $<$ 5) or (24 $<=$ ZFLG $<$ 25) & \citet{2018AA...609A..84S} \\
  COSMOS-Z-COSMOS & 985 & ((4 $<=$ CC $<$ 5) or (24 $<=$ CC $<$ 25)) and (REDSHIFT $>=$ 0.0002) & \citet{2007ApJS..172...70L,2009ApJS..184..218L}\\
  VVDS & 829 & ZFLAGS $==$ 4 or 24 & \citet{2013AA...559A..14L} \\
  DEEP2 & 540 & (ZBEST $>$ 0.001) and (ZERR $>$ 0.0) and (ZQUALITY $==$ 4) and (CLASS $==$ GALAXY) & \citet{2013ApJS..208....5N} \\
  COSMOS-DEIMOS & 517 & (REMARKS $!=$ STAR) and (QF $<$ 10) and (Q $>=$ 1.6) & \citet{2018ApJ...858...77H} \\
  COMOS-Magellan & 183 & (CLASS $==$ nl or a or nla) and (Z\_CONF $==$ 4) & \citet{2009ApJ...696.1195T} \\
  C3R2-Keck & 88 & (REDSHIFT $>=$ 0.001) and (REDSHIFT\_QUALITY $==$ 4) & \citet{2017ApJ...841..111M,2019ApJ...877...81M} \\
  MUSE-Wide & 3 & No filtering conditions. & \citet{2019AA...624A.141U} \\
  UVUDF & 2 & Spectroscopic samples. & \citet{2015AJ....150...31R} \\
\tableline\tableline
\end{tabular}
\end{table*}

\begin{table*}
\centering
\caption{Spectroscopic QSO samples. \label{tab:lood_quasar_samples}}
\hspace*{-1.5cm}\begin{tabular}{c|c p{6cm} c}
\tableline\tableline
Dataset name & Number of objects & Selection conditions & Reference \\
\tableline\tableline
  SDSS DR15 & 290255 & (CLASS $==$ QSO) and (ZWARNING $==$ 0) and (Z\_ERR $>$ 0.0) & \citet{2019ApJS..240...23A} \\
  LAMOST DR5 & 32793 & (CLASS $==$ QSO) and (Z $>$ -9000) & \citet{2012RAA....12.1197C} \\
  OzDES & 772 & (TYPES $==$ AGN or QSO) and (FLAG $!=$ 3 and 6) and (Z $>=$ 0.0025) & \citet{2017MNRAS.472..273C} \\
  PRIMUS & 155 & (CLASS $==$ AGN) and (ZQUALITY $==$ 4) & \citet{2013ApJ...767..118C} \\
  COMOS-Magellan & 53 & (CLASS $==$ bl or bnl or bal) and (Z\_CONF $==$ 4) & \citet{2009ApJ...696.1195T} \\
  6dFGS & 49 & (QUALITY\_CODE $==$ 6) or (REDSHIFT $>$ 1.0) & \citet{2009MNRAS.399..683J} \\
  COSMOS-DEIMOS & 30 & (QF $>=$ 10) and (Q $>=$ 1.6) & \citet{2018ApJ...858...77H} \\
  VVDS & 16 & ZFLAGS $==$ 14 or 214 & \citet{2013AA...559A..14L} \\
  COSMOS-Z-COSMOS & 5 & (CC $==$ 14 or 214) and (REDSHIFT $>=$ 0.0002) & \citet{2007ApJS..172...70L,2009ApJS..184..218L}\\
\tableline\tableline
\end{tabular}
\end{table*}

\begin{table*}
\centering
\caption{Spectroscopic star samples. \label{tab:lood_star_samples}}
  \begin{tabular}{c|c p{6cm} c}
\tableline\tableline
Dataset name & Number of objects & Selection conditions & Reference \\
\tableline\tableline
  LAMOST DR5 & 4131528 & (CLASS $==$ STAR) and (Z $>$ -9000) & \citet{2012RAA....12.1197C} \\
  SDSS DR15 & 544028 & (CLASS $==$ STAR) and (ZWARNING $==$ 0) and (Z\_ERR $>$ 0.0) & \citet{2019ApJS..240...23A} \\
  PRIMUS & 1730 & ZQUALITY $==$ -1 & \citet{2013ApJ...767..118C} \\
  OzDES & 1138 & FLAG $==$ 6 & \citet{2017MNRAS.472..273C} \\
  COSMOS-DEIMOS & 372 & (REMARKS $==$ STAR) and (Q $>=$ 1.6) & \citet{2018ApJ...858...77H} \\
  COSMOS-Z-COSMOS & 300 & REDSHIFT $<$ 0.0002 & \citet{2007ApJS..172...70L,2009ApJS..184..218L}\\
  C3R2-Keck & 1 & (REDSHIFT $<$ 0.001) and (REDSHIFT\_QUALITY $==$ 4) & \citet{2017ApJ...841..111M,2019ApJ...877...81M} \\
\tableline\tableline
\end{tabular}
\end{table*}

As described in Paper I, we use the photometric data retrieved from 
the public data release 1 of the Pan-STARRS1 (PS1) survey 
as input \citep{2010SPIE.7733E..0EK}. The PS1 survey provides 
photometry in five {\it grizy} bands \citep{2016arXiv161205560C}. The input 
data comprise seventeen color-related features: four colors $(g - r)$, $(r - i)$, 
$(i - z)$, and $(z - y)$ in PSF measurement, their uncertainties derived using the 
quadrature rule, the same quantities in Kron measurement, and reddening $E(B - V)$. 
We use only valid photometric data, which are found under 
the condition of {\it ObjectQualityFlags} $==$ {\it QF\_OBJ\_GOOD} 
\citep{2020ApJS..251....7F}. 
To make each input feature contribute 
a similar amount of influence to a model loss function, we rescale input 
features using min-max normalization in a feature-wise manner. 

We use three datasets in training and validating 
our model: the ID, UL, and LOOD datasets. 
Each dataset is used for a different purpose. 
The ID data are galaxy samples used as training data 
for photometric redshift estimation. 
We use 1,480,262 galaxies, which are identical to the samples used in 
Paper I, as summarized in Table \ref{tab:spec_z_samples}. The 
dominant fraction of the training samples comes from the SDSS dataset where 
about half of them are brighter than 19 magnitude in $r$-band 
(see Appendix \ref{sec:append}). 
We assign 80\%, 10\%, and 10\% of the samples 
to the training, validation, and test sets \citep{Murphy12} 
in estimating photometric redshifts for training models, finding 
the best model, and examining the found model, respectively. 
Because we randomly split data and 
have plenty of samples, it is reasonable to assume that the samples 
allocated to each set follow the same distribution as the ID samples. 

The UL data contain UL samples presumably containing 
both ID and OOD samples, and we do not know their 
physical classes and spectroscopic redshifts. 
We use the UL data for the unsupervised training of 
the model for OOD detection. 
We construct the entire UL dataset to have 300,055,711 unknown 
objects following the same selection condition 
adopted for the ID training samples. 
From the UL dataset, we evenly draw samples according 
to right ascension (RA) for unbiased training of the model for OOD detection. 
We use an RA interval of 10\degree~to divide the UL dataset and randomly choose 
1,000,000 samples per RA interval. Hence, 36,000,000 UL samples 
are drawn from the entire UL dataset. 
Then, we assign 80\%, 10\%, and 10\% of the samples to the training, validation, 
and test sets for the purpose of training models, finding optimized models, and 
investigating the found models as described later, respectively.

The LOOD data comprise physically OOD objects; 
we define labeled (i.e., spectroscopically classified) 
QSOs and stars as the LOOD data. 
We obtain the photometric data of 324,234 QSOs and 4,681,989 stars 
in the PS1 DR1 following the same selection condition 
adopted for the ID training samples. 
The sources of these spectroscopic objects are summarized 
in Tables \ref{tab:lood_quasar_samples} and \ref{tab:lood_star_samples}. 
We use the LOOD data for the quantitative assessment of OOD detection performance 
and exclude them from training the models. 
Hence, we use the entire LOOD samples as the test data for 
model validation.

\section{Method \label{sec:method}}

To equip the networks with OOD scoring/detection functionality, we train two neural 
networks $F_{1}$ and $F_{2}$ sharing the same structure in a multi-stage-training 
approach comprising supervised and unsupervised steps (hereinafter, 
$step_{sup}$ and $step_{unsup}$, respectively) 
(see Figure \ref{fig:training_schematics}; \citet{Yu_2019_ICCV}). 
The two networks are trained differently in $step_{unsup}$ to diverge 
their results on potential OOD samples and evaluate OOD score based on their 
inference difference. For UL data, one network tries to have a peaked probabilistic 
distribution of inference on photometric redshifts while the other network intends 
to generate flat probabilistic distribution as explained later in Section 
\ref{subsec:ood_method}. The ID data keep the two networks from diverging on ID 
samples since the training loss using ID samples fosters the two 
networks to have consistent peaked distributions around the true class, i.e., 
correct redshift bin.

The training approach consists of three stages: 
supervised pre-training (training stage-1) with the labeled 
spectroscopic training samples, iterative supervised 
and unsupervised training for OOD scoring/detection (training stage-2) 
with the labeled spectroscopic samples and UL data, respectively, 
and 
supervised training for photometric redshift estimation (training stage-3) 
as the stage-1 conducts. 
The first two stages of the training procedures are originally 
proposed in \cite{Yu_2019_ICCV}, whereas the third stage is added to improve 
the model performance for the original task, i.e., photometric redshift 
estimation, with the added networks for the OOD detection.

We adopt the same supervised learning technique as Paper I in estimating 
photometric redshifts as $step_{sup}$ and the unsupervised method using 
two networks introduced by \cite{Yu_2019_ICCV} in OOD detection as 
$step_{unsup}$. In the following subsections, we outline 
$step_{sup}$, $step_{unsup}$, the three-stage training procedure, 
and assessment metrics to measure model performance. For a more 
detailed explanation of each training step, refer to Paper I and 
\cite{Yu_2019_ICCV}.

\subsection{Supervised Training Step for Photometric Redshift Estimation \label{subsec:photoz_method}}

In $step_{sup}$ using only ID data, we bin 
redshift ranges and classify samples into the redshift bins 
instead of conducting typical regression of redshifts 
as we name the model the multiple-bin regression with the neural network used in Paper I. 
We adopt 64 independent bins with a constant width following the 
performance result in Paper I. Then, the model 
estimates probabilities that the photometric redshifts of input 
samples lie in each bin. Using the model output probabilities, we may 
obtain point-estimation of photometric redshift $z_{phot}$ by averaging 
the central redshift values of the bins with the output probabilities.

As a training loss of $step_{sup}$, we use anchor loss 
$L_{ANCH}$. The loss is designed to 
measure the difference between two given probability distributions 
considering prediction difficulties due to 
various reasons, e.g., the lack of data or the similarities among 
samples belonging to different classes, i.e., redshift bins. 
The anchor loss assigns the difficult 
samples for prediction high weights governed by a weighting parameter $\gamma$. 
For an in-depth definition of the loss, refer to 
\cite{9009013}. Since we train two networks with the same architecture for 
the dual tasks, the training 
loss of $step_{sup}$, which should be {\it minimized}, is set as follow:
\begin{equation}
L_{sup} = L_{ANCH}(p_{1}({\bf z}|{\bf x_{ID}})) + L_{ANCH}(p_{2}({\bf z}|{\bf x_{ID}})),
\end{equation}
where $p_{1}$ and $p_{2}$ are the model output probability 
distributions coming from $F_{1}$ and $F_{2}$, respectively, 
${\bf z}$ is a redshift bin vector, and ${\bf x_{ID}}$ 
is an ID input vector.

\subsection{Unsupervised Training Step for Out-of-Distribution Detection \label{subsec:ood_method}}

$step_{unsup}$ uses the UL data, presumably containing both ID and OOD 
samples, in training models and employs the disparity of the results from 
two networks to evaluate the OOD score of the samples. The two networks 
are trained to maximize disagreement between them in the 
photometric redshift estimation for the potential OOD samples, pushing the OOD 
samples outside the manifold of the ID samples. The 
two networks, which are identically structured and trained on the same ID data 
via supervised learning, 
might have different results on OOD samples because of stochastic effects, although the 
networks are not optimized for OOD detection. Then, by defining discrepancy loss 
$L_{DCP}$ to measure the disagreement and {\it maximizing} it, the two 
networks can be forced to produce divergent results for the potential 
OOD samples, which generally correspond to the samples in the class (i.e., 
redshift bin) boundaries and/or are misclassified (i.e., assigned to an incorrect  
redshift bin). The $L_{DCP}$ defined for this purpose is as follows:
\begin{equation}
    L_{DCP}(p_{1}, \, p_{2}) = H(p_{1}({\bf z}|{\bf x})) - H(p_{2}({\bf z}|{\bf x})),
\end{equation}
where $H(\cdot)$ is the entropy for given 
probability distributions. As the networks are trained to 
maximize the discrepancy $L_{DCP}$, the entropies 
from $F_{1}$ and $F_{2}$ output 
increase and decrease, respectively. The network 
$F_{1}$ outputs a flat probability distribution 
(i.e., high entropy; high uncertainty), 
and the network $F_{2}$ generates a peaked one 
(i.e., low entropy; low uncertainty). 
Notably, the $L_{DCP}$ is available 
in the unsupervised approach using the UL 
samples since it can be computed without sample labels (i.e., redshifts). 
The potential ID samples make it difficult for the two networks produce 
discrepant inference results for photometric redshifts since the inference 
for the ID samples has a strongly peaked distribution in both networks $F_{1}$ 
and $F_{2}$ as explained below. Meanwhile, the potentially OOD samples 
can increase $L_{DCP}$ as the two networks become easily divergent in the training 
process.

To prevent divergent results on the ID samples, we employ the 
sum of $L_{sup}$ and $L_{DCP}$ as the training loss of 
$step_{unsup}$. Although the unsupervised approach using the $L_{DCP}$ 
diverge the model outputs on the probable OOD examples included 
in the UL dataset, it can also cause discrepant 
results on the ID samples since the UL data includes ID and OOD 
samples. Hence, defining the training loss of $step_{unsup}$ as the sum 
help the two network outputs on the ID samples stay as similar as possible 
since minimizing $L_{sup}$ encourages the network outputs 
unimodal probability distributions with a peak at the true class on ID samples. 
Namely, $step_{unsup}$ diverges model outputs for OOD samples more than 
ID ones because OOD samples are out of support of the $L_{sup}$. 
The training loss of $step_{unsup}$, which should be {\it minimized} in training, 
is defined as follows:
\begin{equation}
\label{eq:L_unsup}
 L_{unsup} = L_{sup} + \text{max}(m \,-\, L_{DCP}(p_{1}({\bf z}|{\bf x_{UL}}), \, p_{2}({\bf z}|{\bf x_{UL}})), \, 0), 
\end{equation}
where ${\bf x_{UL}}$ is an UL input vector and $m$ is a margin. Notably, 
$L_{DCP}$ is {\it maximized} as $L_{unsup}$ is {\it minimized} due to the 
negative sign. The use of the margin $m$ is required to prevent overfitting of the networks 
by replacing the second term of the unsupervised loss with zero if $L_{DCP} > m$. 
We use $m = 4.3$ found as the best parameter in multiple trial runs.

Our usage of $L_{unsup}$ as the combination of $L_{sup}$ and 
$L_{DCP}$ can be understood as {\it multi-task learning} of both 
estimating photometric redshifts and evaluating OOD scores. Particularly, 
this multi-task learning model is a case of transferring the knowledge 
of supervised learning (i.e., the model with $L_{sup}$ in estimating photometric 
redshifts) to unsupervised learning (i.e., the model with $L_{unsup}$ in 
evaluating OOD scores) \citep{pmlr-v70-pentina17a}. 
The combined loss in our dual-task learning adopts the 
parameter $m$, which regularizes the contribution of $L_{DCP}$ to the total 
loss $L_{unsup}$. This is a simple strategy to find balance among 
multiple tasks as commonly used as weight parameters with multiple losses 
in {\it multi-task learning} \citep{NEURIPS2018_432aca3a,
8848395,9336293}.

\subsection{Three-Stage Training \label{subsec:three_stage_training}}

\begin{figure*}
\centering
\includegraphics[width=1.\textwidth]{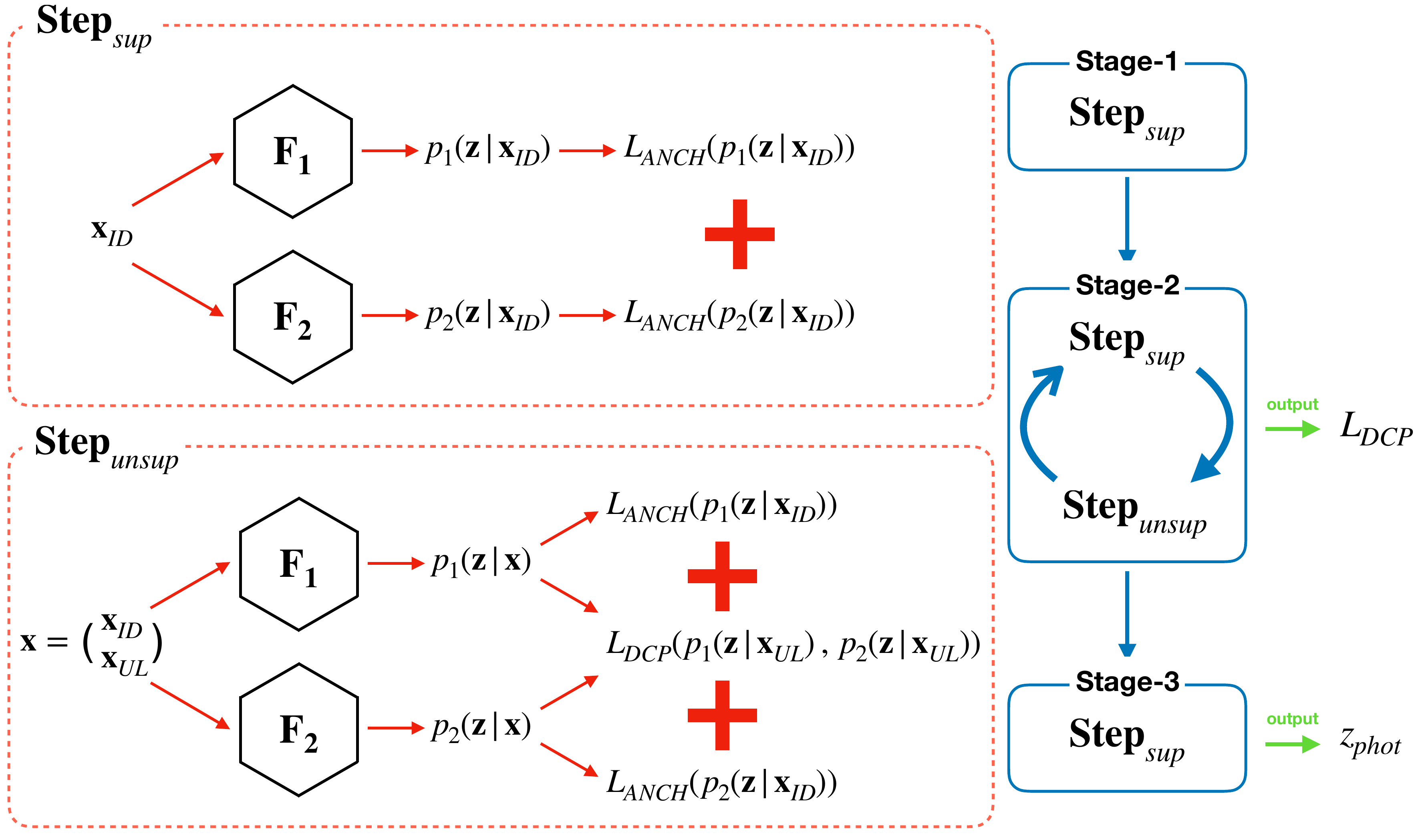}
\caption{Schematic of a three-stage training. 
The subscripts ID and UL of {\bf x} indicate that {\bf x} belongs
 to the ID and UL datasets, respectively. The training stages comprise two 
  learning steps: supervised step ($step_{sup}$; {\it upper 
 red box}) and unsupervised step ($step_{unsup}$; {\it bottom red box}). $step_{sup}$ is designed for photometric 
 redshift estimation, and $step_{unsup}$ is contrived for OOD scoring/detection. Notably, the type of loss for each 
 step differs. As the first stage of the entire training procedure (training stage-1), the networks are 
 pre-trained with $step_{sup}$. Then, the pre-trained networks are forwarded to training stage-2 and trained for OOD scoring/
 detection using iterative $step_{sup}$ and $step_{unsup}$. In the training stage-3, the networks are again optimized for 
 photometric redshift estimation using only $step_{sup}$. \label{fig:training_schematics}}
\end{figure*}

We proceed with the three-stage training of the two networks. 
Each training stage comprises aforementioned $step_{sup}$ and $step_{unsup}$ 
as summarized in Figure \ref{fig:training_schematics}. 
Training stage-1 is a pre-training of the two networks with the $step_{sup}$. The purpose of this stage is to 
obtain a stable performance of the networks in estimating photometric redshifts. 
In training stage-2, the two networks are trained for OOD detection 
using an iterative training procedure comprising one $step_{sup}$ and 
two $step_{unsup}$s. Although $L_{sup}$ included in $L_{unsup}$ helps the 
disparity between two network outputs on ID samples stay small, the disparity 
increases as the models are trained by the optimization using UL samples and $L_{DCP}$. 
Composing the stage-2 as the iterative training procedure incorporating $step_{sup}$ 
further precludes the disparity on the ID samples from increasing as $L_{sup}$ included 
in $L_{unsup}$ affects the total loss (see Equation \ref{eq:L_unsup}). 
Therefore, as the training continues, the disagreement of the network outcomes 
for the probable OOD samples becomes larger than that for the ID samples because 
the OOD samples are outside the support of the $step_{sup}$, which is only applied to 
the ID galaxies. Then, the measure of disagreement (i.e., $L_{DCP}$) can be used as 
an OOD score to flag OOD candidates. At the inference level, the trained networks in 
training stage-2 are used to score the test samples as OOD.

In addition to training stage-1 and -2, we proceed to an additional supervised training 
stage, training stage-3, with $step_{sup}$ for photometric redshift estimation. 
The unsupervised training for OOD detection affects the model 
performance of the original task (i.e., photometric redshift estimation) 
because the model in training stage-2 is optimized for two 
losses described in Equation \ref{eq:L_unsup}. In practice, training stage-3 might 
not be needed if the models trained in stage-1 are saved and used 
for the task of photometric redshift estimation. 
In this work, we train the two networks again 
using only $L_{sup}$ after training stage-2. Then, the outputs of the two stage-3 networks 
are combined by averaging them to estimate photometric redshifts.

\subsection{Assessment Metrics}

Since our model performs the two tasks together, 
we need separate metrics to assess model 
performances on both tasks. As point-estimation metrics 
in estimating photometric redshifts with photometric 
redshifts as mean value with probabilistic inference by the 
trained model, we 
adopt the ones used in Paper I: bias, MAD, $\sigma$, 
$\sigma_{68}$, NMAD, and $R_{cat}$. For a detailed explanation 
of the metrics, refer to Paper I. 
In short, the bias is measured as the absolute mean of $\Delta z$ which 
is difference between photometric redshifts and true spectroscopic redshifts. 
Similarly, the MAD is the mean of the absolute difference $|\Delta z|$. 
$\sigma$ and $\sigma_{68}$ corresponds 
to the standard deviation of $\Delta z$ and the 
68th percentile of $|\Delta z|$, respectively. 
NMAD is 
defined to be 1.4826 $\times$ median of $\Delta z$. $R_{cat}$ 
presents the fraction of data with $|\Delta z| > 0.15$. 
Notably, the lower 
the point-estimation metrics are, the higher the performance is.

Before explaining OOD detection metrics, we first define four 
measures: true negative (TN), true positive (TP), false negative (FN), 
and false positive (FP). The Boolean values (i.e., true 
and false) in the terms indicate whether the predicted class of the 
sample is correct or incorrect. The negative and positive in the terms 
mean the true class in which the given sample is included. We, respectively, 
set ID and OOD as negative and positive since the model performs OOD detection 
as a task. Hence, for example, the number of true positive cases is 
the number of correctly classified OOD samples. The numbers of 
these four cases can be 
varied with respect to the adopted threshold of the OOD score. Notably, 
these metrics are used with the LOOD samples not in training the model 
but in validating the trained model.

We use the central value between the minimum and maximum $L_{DCP}$ as the 
threshold to compute the metrics. Note that this threshold is chosen simply because 
it can be easily estimated for model assessment. 
The following is a brief explanation of the detection metrics.
\begin{itemize}
    \item {\it Accuracy}: ratio of the correctly classified samples 
    to the entire samples $\frac{{\it TP}+{\it TN}}{{\it TP}+{\it TN}+{\it FN}+{\it FP}}$,
    \item {\it Precision}: ratio of the 
    correctly classified positive samples to the entire positively 
    classified samples $\frac{{\it TP}}{{\it TP} + {\it FP}}$,
    \item {\it True Positive Rate (TPR)} or {\it recall}: fraction 
    between the correctly predicted positive samples and the entire 
    positive samples $\frac{{\it TP}}{{\it TP} + {\it FN}}$,
    \item {\it $F_{\beta}$}: harmonic mean of 
    precision and recall weighted by $\beta$ $\frac{(1+\beta^{2}) * 
    Precision * Recall}{\beta^{2} * Precision + Recall}$. The lesser 
    and larger $\beta$ gives more weight to precision and recall, 
    respectively. Typical values for $\beta$ is 0.5, 1.0, and 1.5. In 
    this work, we set $\beta = 1.5$ ($F_{1.5}$) to 
    focus on more recall than precision, paying attention to the number of the 
    correctly classified OOD samples among all OOD samples,
    \item {\it $AUC_{ROC}$}: area under the receiver operating 
      characteristic (ROC) curve. The ROC curve is a visualized 
    measure of the detection performance for the binary classifier. In 
    the curve, the {\it TPR} is plotted as a function of the false positive 
    rate for different thresholds of the OOD score. False 
    Positive Rate is a fraction between the incorrectly predicted 
    negative samples and entire negative samples $\frac{{\it FP}}{{\it TN} + 
    {\it FP}}$. The {\it $AUC_{ROC}$} offers a comprehensive measure of the 
    detection performance of the model.
  \item {\it $AUC_{PR}$}: area under the precision-recall (PR) curve. 
    Although $AUC_{ROC}$ offers a representative measure of 
    binary detection performance generally, it may 
    produce misleading results for imbalanced data because {\it TPR} 
    (i.e., the vertical axis of the ROC curve) depends only on positive cases 
    without considering negative cases. In the PR curve, 
    precision is plotted as a function of recall. Because precision 
    considers both positive and negative samples, $AUC_{PR}$ may 
    be more informative for imbalanced data.
\end{itemize}
The higher values of the detection metrics indicate better detection performance 
of the model. Notably, these OOD detection metrics are evaluated with the 
labeled ID (i.e., galaxies) and OOD (i.e., stars and QSOs confirmed 
spectroscopically).

\section{Results \label{sec:result}}
We present the results of the model in testing 
the two tasks. All results in this section are produced 
with the samples in the test dataset unless otherwise stated.

\subsection{Photometric Redshift Estimation on In-Distribution Samples \label{subsec:phot_red}}
\begin{table*}
\begin{center}
\caption{Metrics for photometric redshift estimation of the models from training stage-2 and -3. \label{tab:photo_red_metrics}}
\hspace*{-2.35cm}\begin{tabular}{c|c c c c c c}
\tableline\tableline
Variables & \multicolumn{6}{c}{metrics} \\
\cline{1-7}
Case & Bias & MAR & $\sigma$ & $\sigma_{68}$ & {\it NMAD} & $R_{cat}$ \\
\tableline\tableline
Stage-2 (high entropy) & 0.0306 & 0.0466 & 0.0959 & 0.0343 & 0.0314 & 0.0637 \\
Stage-2 (low entropy) & {\bf 0.0015} & 0.0281 & 0.0444 & 0.0293 & 0.0280 & 0.0150 \\
Stage-3 (average) & 0.0020 & {\bf 0.0254} & {\bf 0.0392} & {\bf 0.0272} & {\bf 0.0256} & {\bf 0.0087} \\
\tableline\tableline
\end{tabular}
\end{center}
\end{table*}

\begin{figure*}
\centering
\includegraphics[width=1.\textwidth]{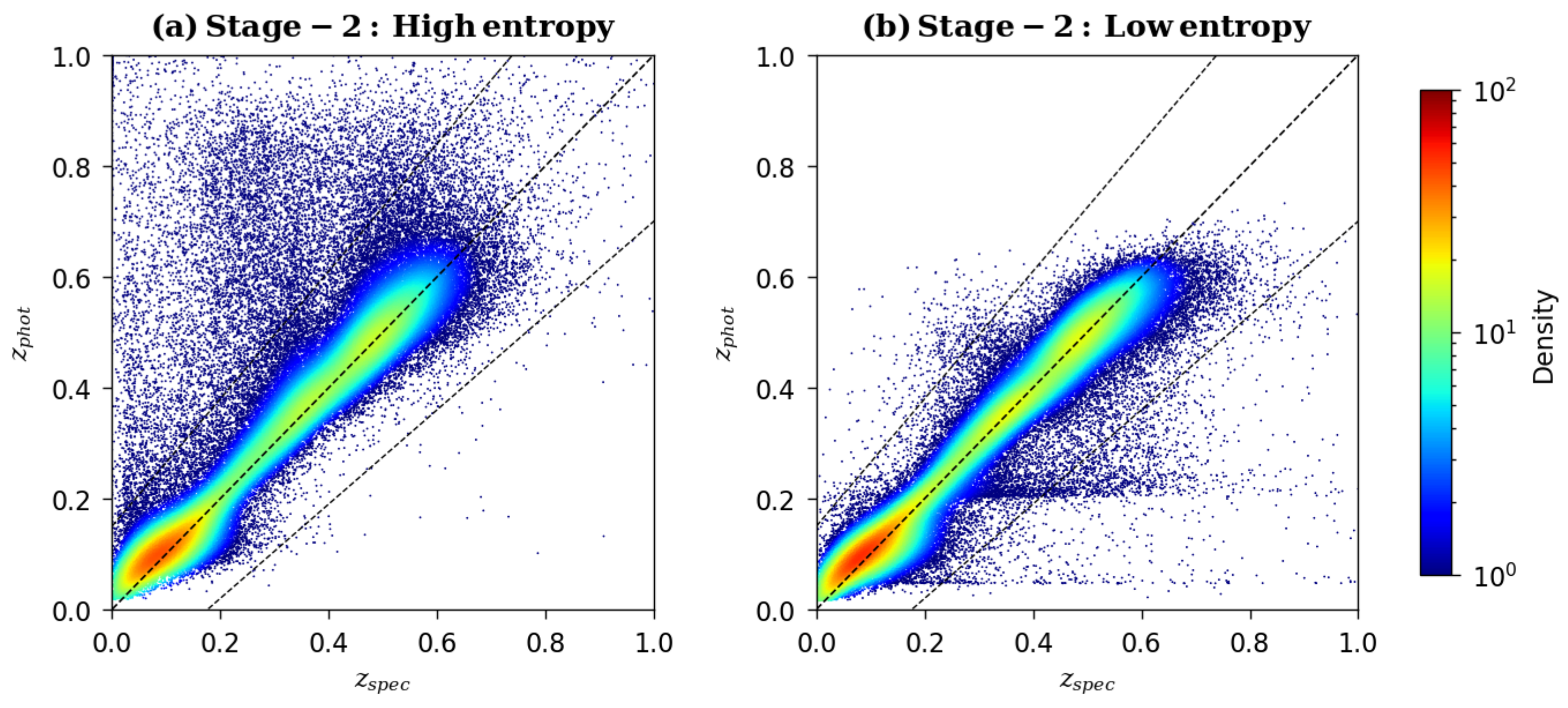}
\caption{Comparison between spectroscopic and photometric redshifts from the high entropy 
  ($F_{1}$; {\it left}) and low entropy ($F_{2}$; {\it right}) networks in training stage-2. 
  The scatters are color-coded according to number density. 
  The dashed lines correspond to slope-one lines and catastrophic error boundaries, 
  respectively. \label{fig:red_scatter_cden_stage2}}
\end{figure*}
\begin{figure}
\centering
\includegraphics[width=.5\textwidth]{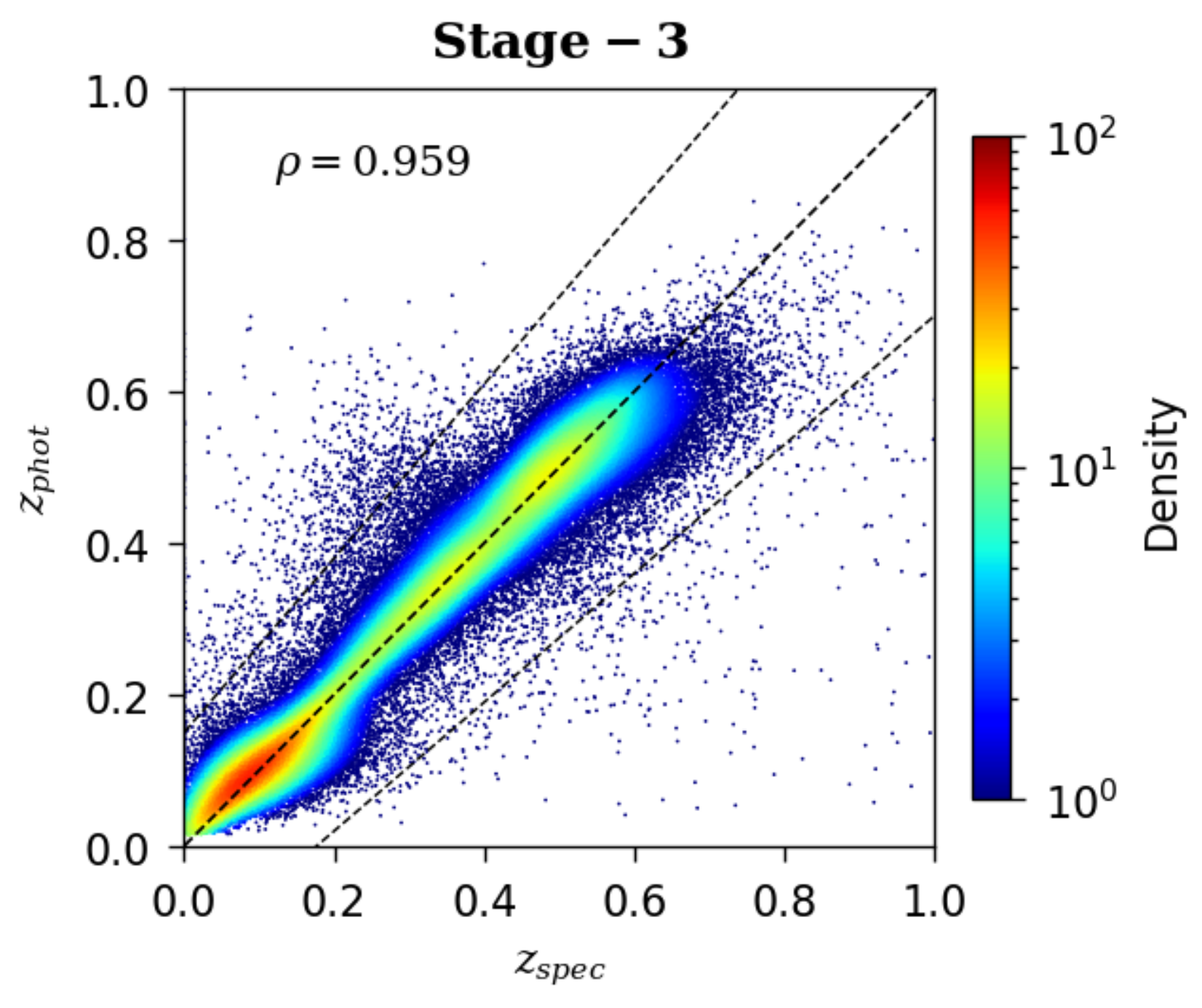}
\caption{Comparison between spectroscopic and photometric redshifts from 
  the averaged network in training stage-3. The lines and color-coding follow 
  Figure \ref{fig:red_scatter_cden_stage2}. 
  The Pearson correlation coefficient $\rho$ 
  proves good agreement between spectroscopic and photometric redshifts.
  Notably, the peculiar patterns found for the models in training stage-2 
  disappear in the stage-3 model. \label{fig:red_scatter_cden_stage3}}
\end{figure}

We find that the networks in training stage-3 outperform the networks in the stage-2 
in terms of photometric redshift estimation. To verify the 
effectiveness of the additional stage training stage-3, 
we compare the quality of the photometric redshifts derived from 
the models in training stage-2 and -3. Table \ref{tab:photo_red_metrics} shows 
the point-estimation metrics evaluated in the high and low entropy models 
(high and low entropy models, i.e., $F_{1}$ and $F_{2}$, respectively) in training stage-2 and 
the averaged model in stage-3. 
Except for the bias, the training stage-3 model outperforms others. 
It proves that unsupervised training 
of the networks in training stage-2 for OOD detection affects the quality of photometric 
redshifts as expected for multi-task learning.

The networks in training stage-2 hold noticeable peculiarities showing 
deteriorated photometric redshifts. 
As shown in Figure \ref{fig:red_scatter_cden_stage2}, 
the photometric redshifts from the high entropy model are 
overestimated for a large fraction of samples. 
Meanwhile, the photometric redshifts from the 
low entropy model show concentrations at $z_{phot} \sim 0.05$ and $z_{phot} \sim 0.2$. 
The peculiarities in the derived photometric redshifts 
arise from the minimization of 
$L_{unsup}$ during training stage-2, which makes the output probability distributions 
of the high and low entropy models flat and peaked, respectively. 
These oddities of the training stage-2 model vanish in the averaged stage-3 model 
(see Figure \ref{fig:red_scatter_cden_stage3}). 
The Pearson correlation coefficient $\rho = 0.959$ affirms that 
the estimated photometric redshifts match spectroscopic 
redshifts well.

\subsection{Out-of-Distribution Score for Labeled Data \label{subsec:id_res}}

\begin{table*}
\begin{center}
\caption{OOD flagging metrics on LOOD samples. Every metric higher than 0.98 
  with the networks in training stage-2 proves that the networks accurately detect OOD objects. \label{tab:detect_metrics}}
\hspace*{-2.35cm}\begin{tabular}{c| c c c c c c}
\tableline\tableline
Step & Accuracy & Precision & Recall & $F_{1.5}$ & $AUC_{ROC}$ & $AUC_{PR}$\\
\tableline\tableline
Stage-2 & {\bf 0.9802} & 0.9987 & {\bf 0.9808} & {\bf 0.9863} & {\bf 0.9938} & {\bf 0.9998} \\
Stage-3 & 0.0295 & {\bf 0.9990} & 0.0008 & 0.0012 & 0.7547 & 0.9914 \\
\tableline\tableline

\end{tabular}
\end{center}
\end{table*}

\begin{figure}
\centering
\includegraphics{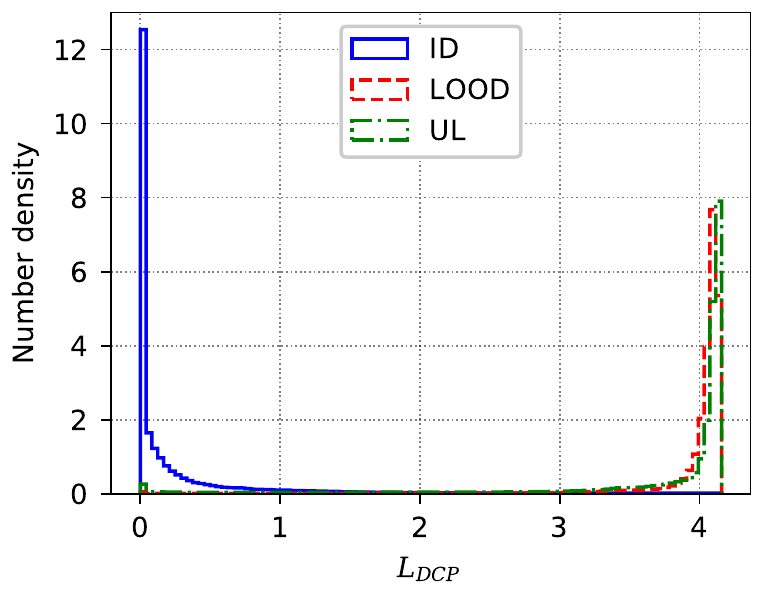}
\caption{$L_{DCP}$ distribution of ID, LOOD, and UL samples. 
$L_{DCP}$ is an OOD score indicating how likely the given 
sample is to be OOD. The samples with high $L_{DCP}$ are more 
liable to be OOD than the ones with low $L_{DCP}$. 
Most of the ID and LOOD samples have low and high $L_{DCP}$, 
respectively. Providing that UL samples are mostly distributed in 
the high $L_{DCP}$ range, we infer that majority of UL samples are 
OOD objects with respect to the trained model for photometric redshifts 
although we cannot classify the samples. \label{fig:dcp_loss_dstrb}}
\end{figure}
\begin{figure*}
\centering
\includegraphics[width=1.\textwidth]{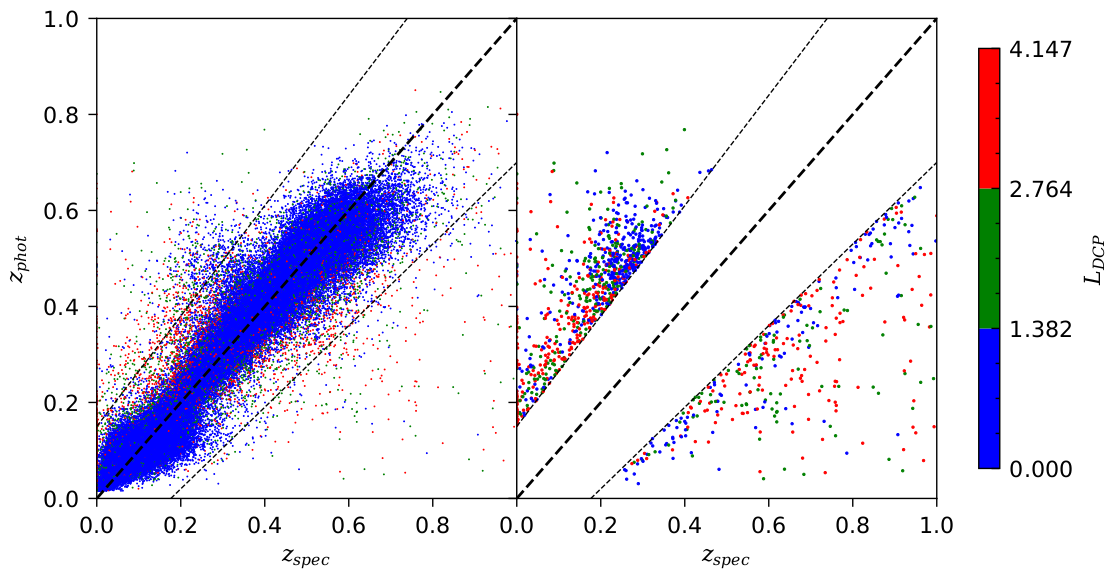}
\caption{Distribution of $L_{DCP}$ in training stage-2 with respect to the 
spectroscopic and photometric redshifts derived from the stage-3 model. 
  The distribution is given for the entire test samples ({\it left}) 
  and the catastrophic samples ({\it right}) separately. 
For clear visualization, we divide 
$L_{DCP}$ range into three parts with uniform width: low, middle, and 
high $L_{DCP}$ ranges. Notably, the samples with high $L_{DCP}$ 
  mainly reside near or outside the catastrophic boundaries (dotted lines). 
\label{fig:red_scatter_cdcp}}
\end{figure*}
\begin{figure*}
\centering
\plottwo{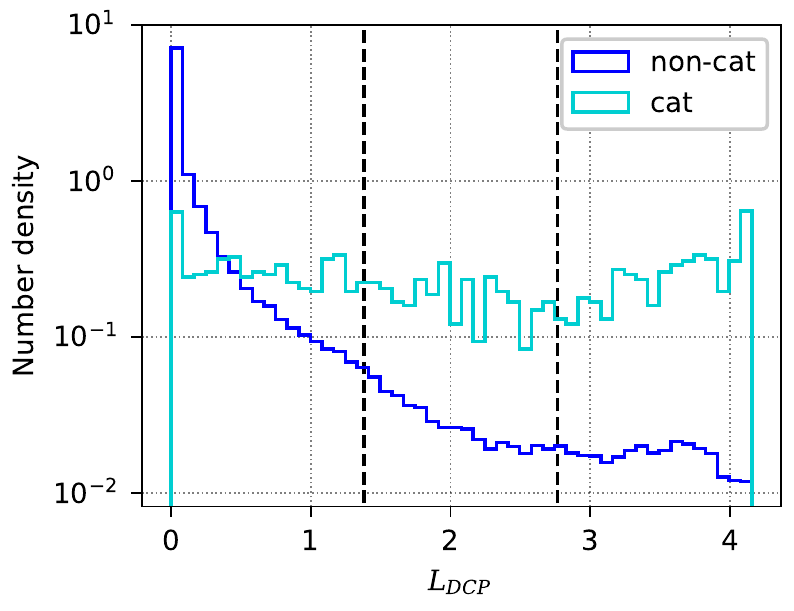}{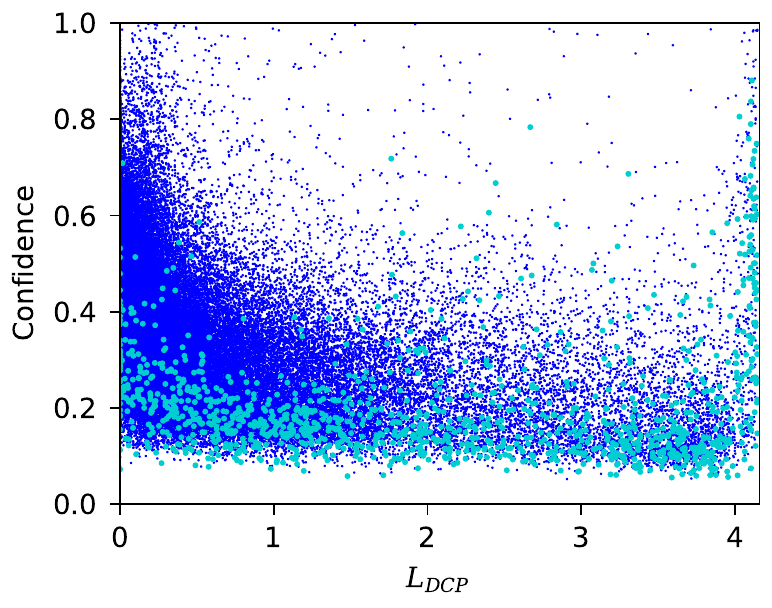}
\caption{{\it Left}: $L_{DCP}$ distributions of ID {\it non-cat} (blue) and 
  {\it cat} (azure) samples. The vertical dashed lines are boundaries of low, 
  middle, and high $L_{DCP}$ ranges. {\it Right}: Distributions of 
  ID {\it cat} and {\it non-cat} samples with respect to $L_{DCP}$ in training 
  stage-2 and confidence of the training stage-3 model (i.e., the peak 
  value of the model output probability distribution). For clear visualization, 
  we use a larger marker size for the {\it cat} samples and plot {\it cat} 
  samples on top of {\it non-cat} ones. \label{fig:dcp_loss_dstrb_id}}
\end{figure*}
\begin{figure*}
\centering
\includegraphics[width=1.\textwidth]{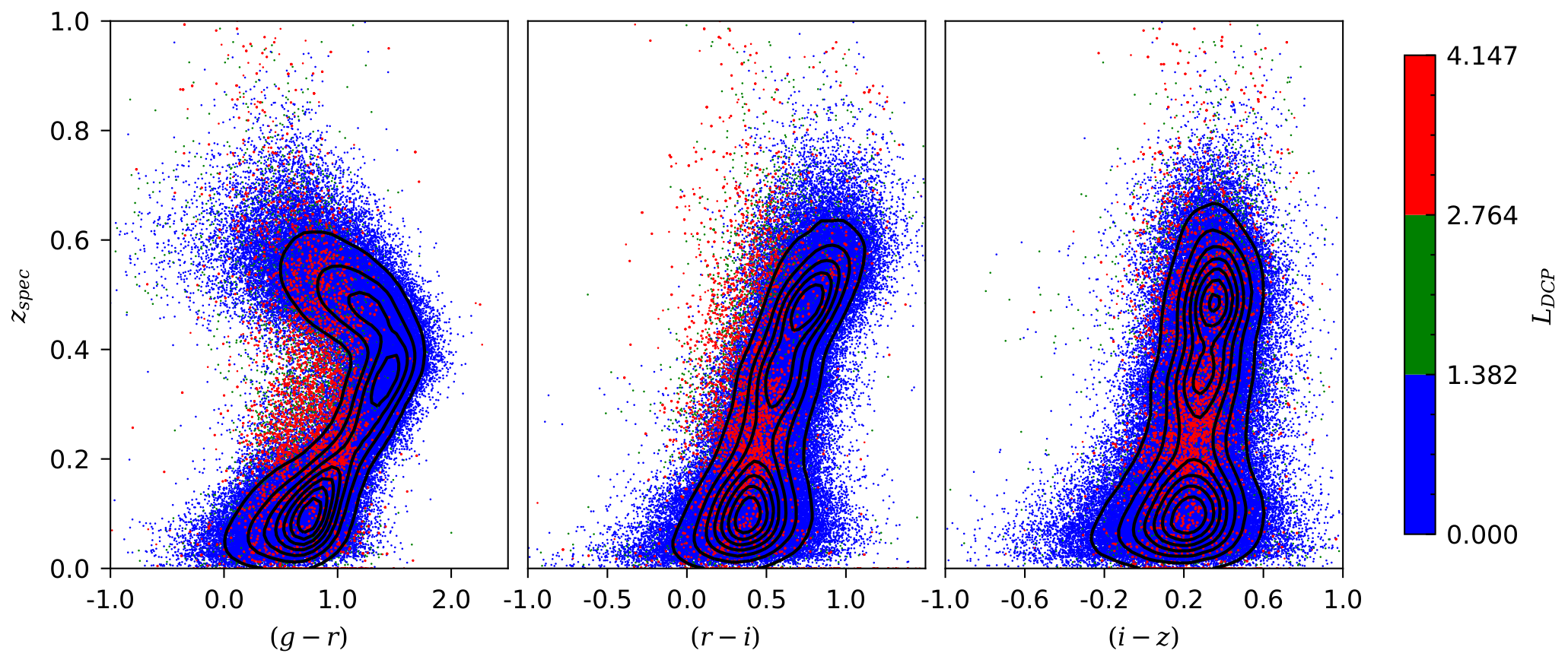}
\caption{$L_{DCP}$ distributions of the ID samples in the space of 
  PSF-measurement colors and spectroscopic redshifts. 
  The contour lines depict the distributions of the training samples in each input space. 
  Notably, high $L_{DCP}$ samples mostly lie outside the contour lines or 
  in the low density regions. \label{fig:red_col_scatter_id}}
\end{figure*}

\begin{figure*}
\centering
\includegraphics{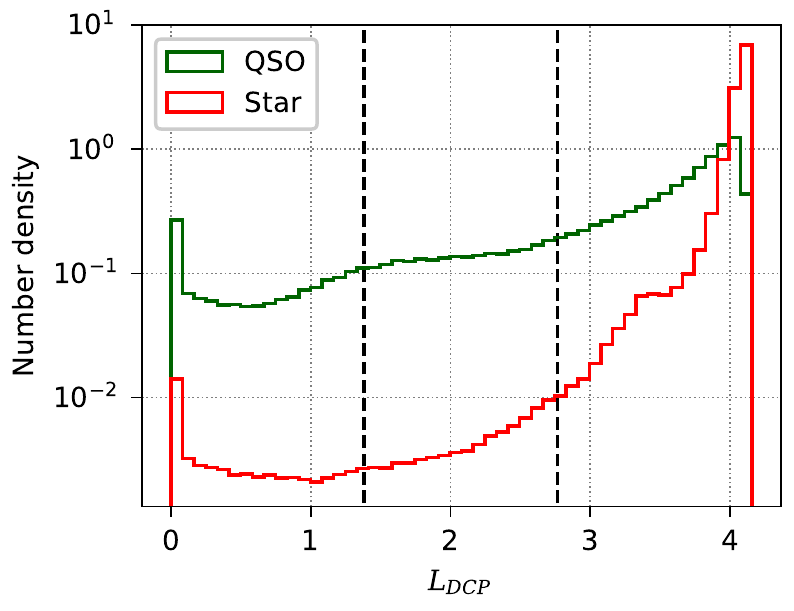}
\caption{$L_{DCP}$ distributions of the LOOD samples: QSOs (darkgreen) and stars (red). 
Vertical dashed lines are boundaries of low, middle, and high $L_{DCP}$ ranges 
  same as the ranges used in Figures \ref{fig:red_scatter_cdcp} and \ref{fig:red_col_scatter_id}. 
  \label{fig:dcp_loss_dstrb_lood}}
\end{figure*}
\begin{figure*}
\centering
\subfigure[QSO]{\includegraphics[width=1.\textwidth]{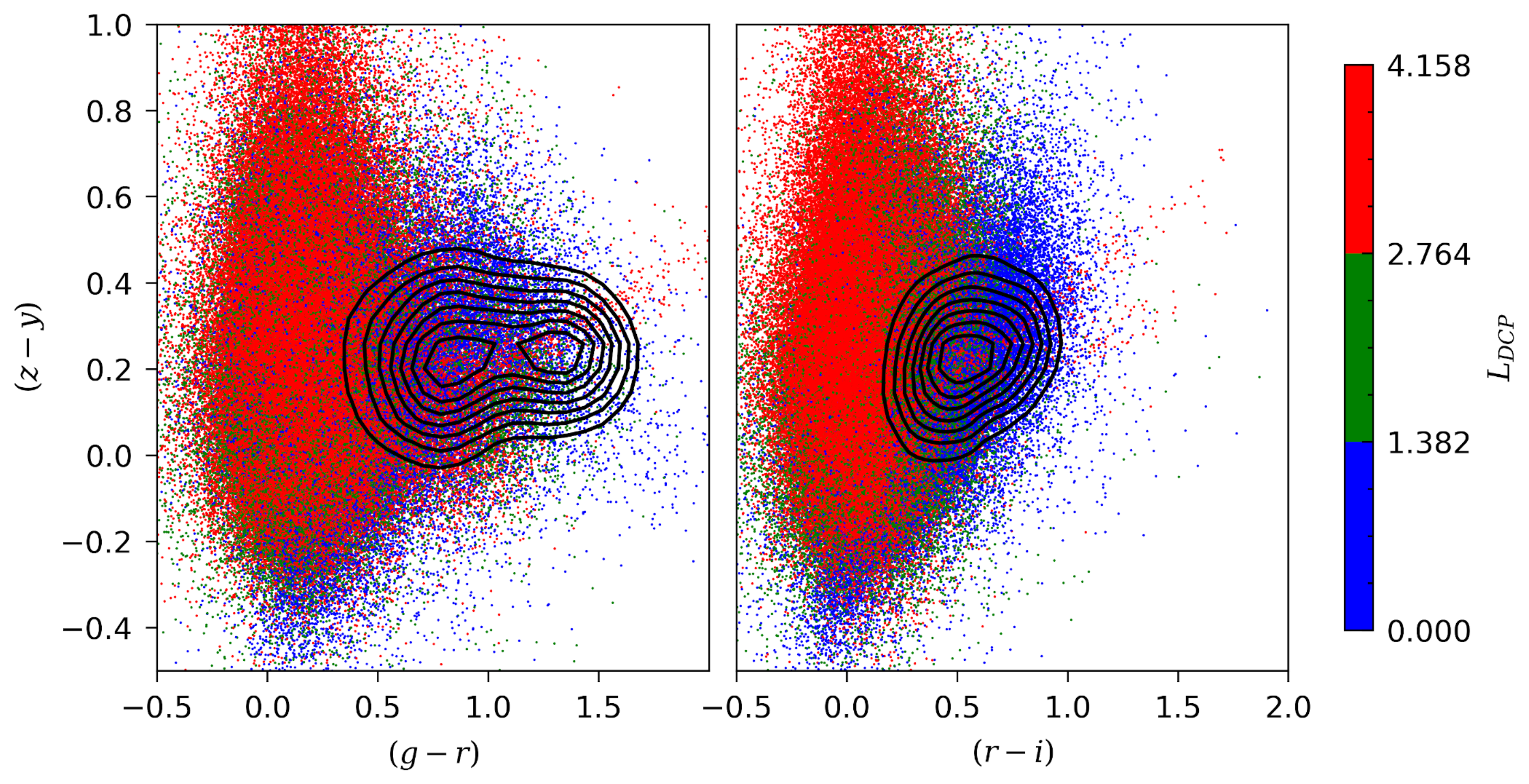}}
\subfigure[Star]{\includegraphics[width=1.\textwidth]{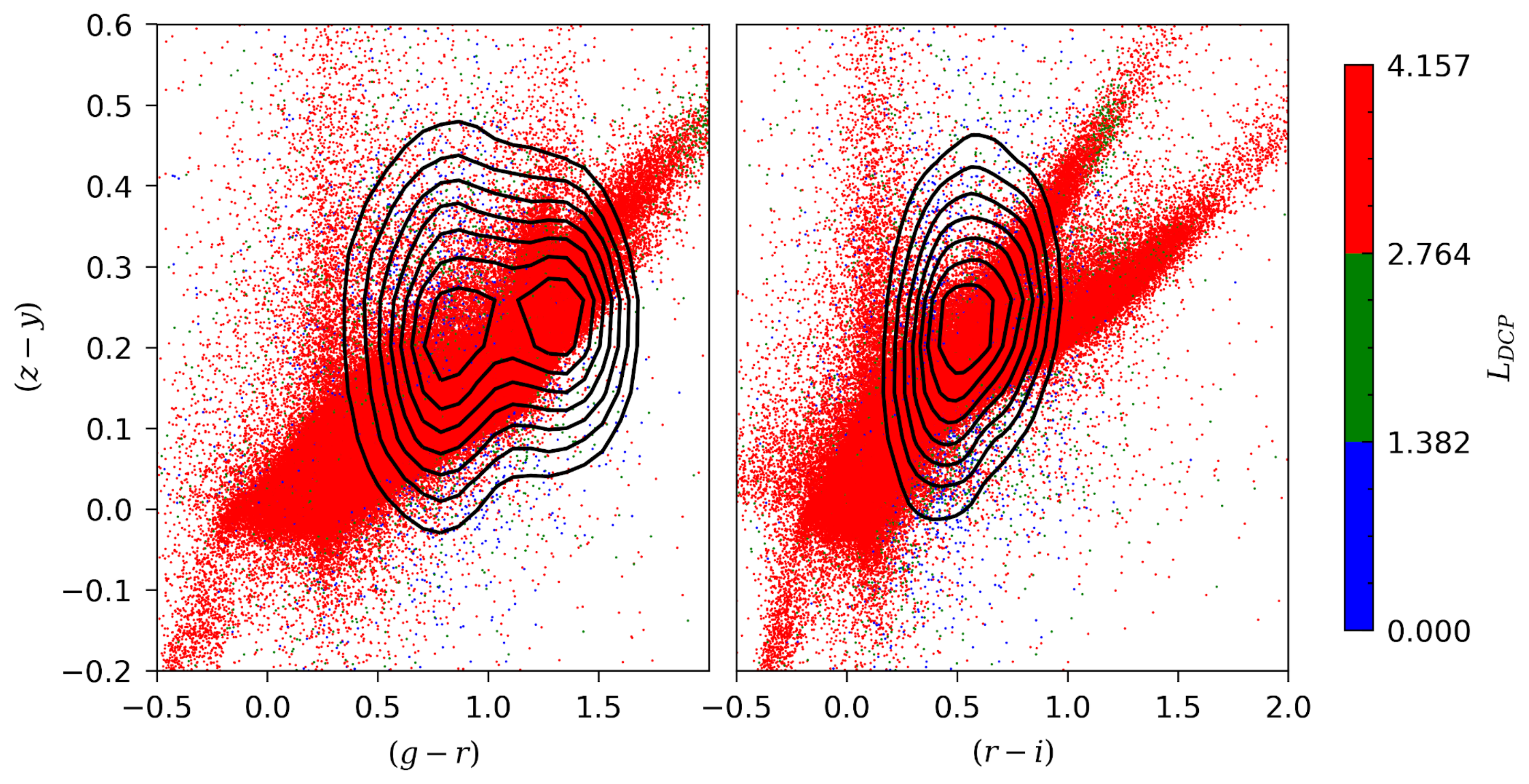}}
\caption{$L_{DCP}$ distributions of QSOs ({\it upper}) and stars ({\it bottom}) 
  with respect to the input colors in PSF measurement. The 
black contour lines depict the ID training sample distribution. The points are 
discretely color-coded according to $L_{DCP}$ groups. 
  The samples of stars used to plot here are 
randomly under-sampled approximately by a factor of 10 from the entire star 
  LOOD dataset. \label{fig:colcolscatter_lood}}
\end{figure*}
\begin{figure*}
\centering
\includegraphics[width=1.\textwidth]{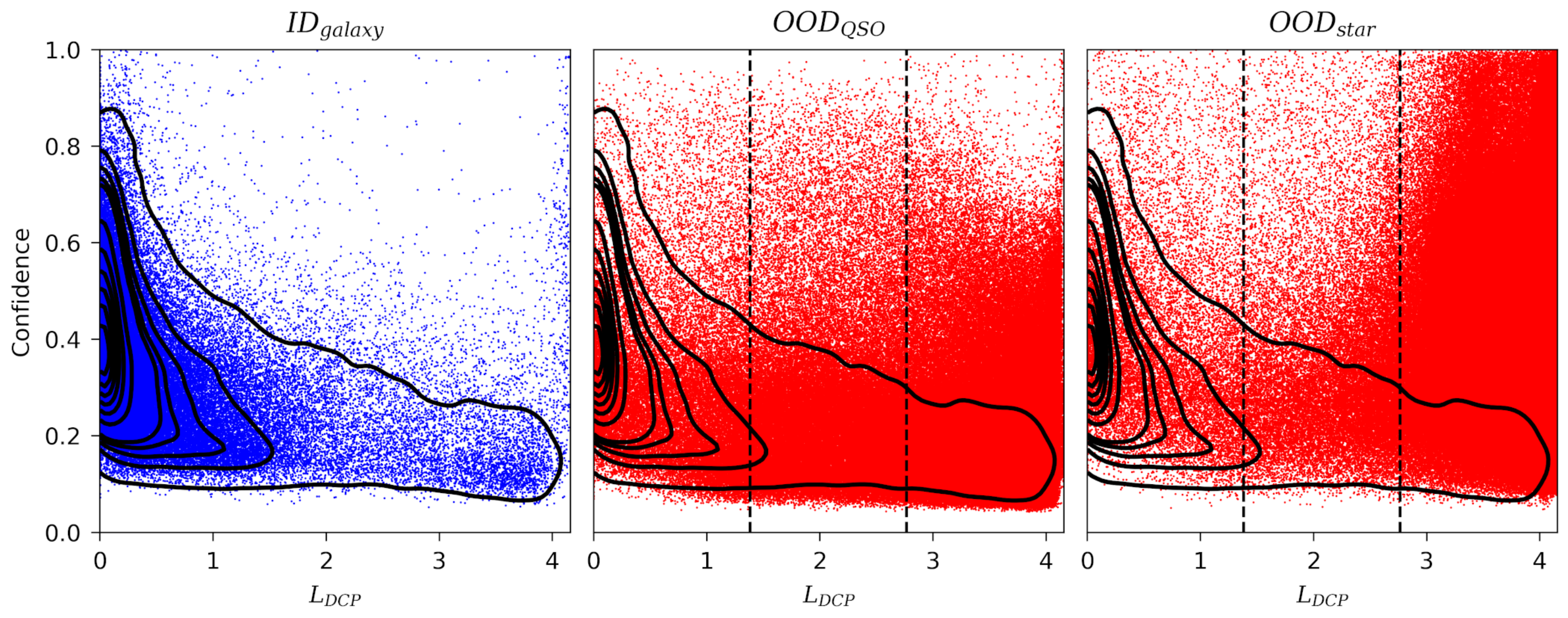}
\caption{Distribution of galaxy (ID, {\it left}), QSO (OOD, {\it middle}), 
  and star (OOD, {\it right}) samples with respect to $L_{DCP}$ and confidence 
  of photometric redshifts (i.e., the peak value 
  of the model output probability distribution) for a given sample. 
  The contour lines indicate the ID sample distribution. 
  The guide values of $L_{DCP}$ dividing low, middle, and high $L_{DCP}$ 
  ranges are presented as vertical dashed lines. \label{fig:dcp_conf_scatt}}
\end{figure*}

As we design and expect, training stage-2 model shows better separation between the ID and OOD samples 
(i.e., physically OOD samples of stars and QSOs) than the stage-3 model. 
We compare the OOD detection performance of the 
networks in training stage-2 and -3. The detection metrics of the networks in training stage-2 
and -3 are tabulated in Table \ref{tab:detect_metrics}, showing 
that the OOD detection model is optimized in training stage-2.

Interestingly, the training stage-3 model has higher precision than the stage-2 model, 
which is due to the high {\it TP} values with the small number 
of samples flagged as OOD by the training stage-3 model; 352,055 samples 
($\sim 7\%$) among 5,006,223 true OODs are correctly classified as OOD. 
This small number results in unexpectedly high precision, $AUC_{ROC}$, and 
$AUC_{PR}$ for the training stage-3 model. 

The stage-2 model separates well the labeled ID and OOD samples 
as shown in Figure \ref{fig:dcp_loss_dstrb} displaying the $L_{DCP}$ 
distributions of the ID, LOOD, and UL samples in training stage-2. 
Although most ID samples appear in low 
$L_{DCP}$ ranges, most LOOD samples are distributed in the 
vicinity of the highest $L_{DCP}$. In addition, most 
UL samples show high $L_{DCP}$ values. 
There are small portions of the UL and LOOD samples with the low $L_{DCP}$, 
which represents samples with the input properties similar to those 
of the ID samples. 
We infer that the highest proportion of UL samples are more likely 
to be OOD objects (including physically OOD objects such as stars and QSOs) 
than ID galaxies although we cannot specify the classes of the 
samples. The distribution of the $L_{DCP}$ in the UL samples makes 
sense because the largest fraction of sources should be 
stars in our Galaxy.

The prediction difficulties of photometric redshift 
estimation and OOD scores for the ID data are related 
as presented in Figure \ref{fig:red_scatter_cdcp}. 
Although most samples with {\it non-catastrophic} 
(hereinafter, non-cat) estimation 
(i.e., $(z_{spec} ~ - ~ z_{phot})/(1 + z_{spec}) \le 0.15$) 
are assigned to the low $L_{DCP}$ group, most samples with 
{\it catastrophic} (hereinafter, cat) estimation of photometric redshifts 
belong to middle or high $L_{DCP}$ groups.  
High $L_{DCP}$ samples generally more deviate from 
the slope-one line than low $L_{DCP}$ samples, indicating that the samples 
with high OOD score can be viewed as OOD-like ID samples and tend to have 
incorrect photometric redshifts.

The distributions of $L_{DCP}$ for the ID {\it non-cat} and 
{\it cat} samples shown in Figure \ref{fig:dcp_loss_dstrb_id} 
also affirm our interpretation of the relation between 
$L_{DCP}$ and the error in photometric redshifts. 
The distributions show that a higher fraction of 
{\it cat} samples populates in the high $L_{DCP}$ ranges than 
{\it non-cat} samples. Although the number of {\it non-cat} 
samples decreases as the $L_{DCP}$ increases, 
the {\it cat} samples are almost uniformly distributed over the entire 
$L_{DCP}$ range. The difficulties in estimating photometric redshifts 
are related to the OOD-like ID samples\footnote{In Paper I, we have 
already found that the lack of training samples 
causes high errors in photometric redshifts.}.

The training stage-3 model is unreliable concerning 
{\it cat} samples with high $L_{DCP}$. 
As shown in Figure \ref{fig:dcp_loss_dstrb_id}, 
the number of {\it non-cat} samples with high confidence 
on photometric redshifts decreases 
as $L_{DCP}$ increases. The {\it cat} samples mostly have low 
confidence on photometric redshifts throughout low and middle $L_{DCP}$ ranges. 
It also shows that the networks in the training stage-3 are well calibrated in 
estimating photometric redshifts for the {\it cat} samples 
with low and middle $L_{DCP}$ presenting the low confidence on them. 
However, the {\it cat} samples with high $L_{DCP}$ 
extend to cases with high confidence which is the 
indication of overconfident results on the wrong 
photometric redshift estimation for OOD-like ID {\it cat} samples. 
It also emphasizes the importance of the OOD 
detection algorithm since we cannot filter the OOD samples simply 
based on confidence on photometric redshifts without the OOD score. 

The distribution of $L_{DCP}$ in the space of input features 
allows us to visually inspect the OOD-like ID samples. 
In Figure \ref{fig:red_col_scatter_id}\footnote{
This visual presentation often used hereafter displays 
the projected dimensions of the input space with degeneracy of 
the input features in the higher dimension.}, most high $L_{DCP}$ 
samples reside in the low-density regions of the training samples. 
The training stage-2 model is optimized to 
assign high OOD scores to 
under-represented galaxies, meaning that these high $L_{DCP}$ ID 
samples are under-represented galaxies from the model perspective. 

As we find the ID samples with high $L_{DCP}$, there are LOOD 
samples with low $L_{DCP}$ (i.e., the ID-like LOOD samples). 
As presented in Figure \ref{fig:dcp_loss_dstrb_lood} depicting the $L_{DCP}$ 
distributions for the two types of LOOD objects, i.e., QSOs and stars, 
a fraction of LOOD samples have low 
$L_{DCP}$ although most samples have high $L_{DCP}$. 
The points to be noted are as follows: 1) both QSO and star distributions have 
peaks at the lowest $L_{DCP}$ bin, 2) the ratio of the lowest $L_{DCP}$ 
peak to the highest $L_{DCP}$ peak for QSO is larger than that of star, 
and 3) more QSOs are distributed in the low and middle 
$L_{DCP}$ ranges than stars.

The ID-like LOOD samples locate around the ID sample distribution in 
the input space. Figure \ref{fig:colcolscatter_lood} shows 
that the QSO samples with high $L_{DCP}$ overwhelm the ones 
with low $L_{DCP}$ outside the 
ID sample density contours. However, the number of the low 
$L_{DCP}$ samples for QSOs increases as they approach the ID 
sample density peak. These ID-like QSOs contribute to the 
peak at the lowest $L_{DCP}$ bin of Figure \ref{fig:dcp_loss_dstrb_lood}.

As found in Figure \ref{fig:colcolscatter_lood}, 
some stars certainly reside inside the contours of the ID 
galaxy samples with respect to the projected input dimension, 
corresponding to the stars with low $L_{DCP}$ 
(Figure \ref{fig:dcp_loss_dstrb_lood}). However, most stars 
have input features significantly deviating from the typical 
properties of the ID galaxy samples in any one dimension of the 
input features. Compared with the QSO distribution shown in Figure 
\ref{fig:colcolscatter_lood}, 
more stars than QSOs can be sorted out from the ID samples in higher 
dimensions of input features since 
the star distribution significantly differs from that of the ID samples. 
Therefore, the relatively smaller peak of stars at the lowest OOD 
score and a lower fraction of ID-like stars than QSOs arise as shown in 
Figure \ref{fig:dcp_loss_dstrb_lood}.

Filtering out the QSO and star samples with low/middle $L_{DCP}$ enables further 
reliable usage of the trained model for photometric redshifts. As depicted 
in Figure \ref{fig:dcp_conf_scatt}, although the confidence of ID samples gradually reduces 
as $L_{DCP}$ increases, the LOOD objects with high $L_{DCP}$ have 
higher confidence than those with low $L_{DCP}$. Using this distribution 
difference in the space of $L_{DCP}$ and confidence, a considerable number 
of LOOD samples with low and middle $L_{DCP}$ can be filtered by 
checking their higher confidence than that of ID samples.

\subsection{Out-of-Distribution Score and Photometric Redshifts of Unlabeled Data}
\label{sec:unlabeld}
\begin{figure}
  \centering
  \includegraphics{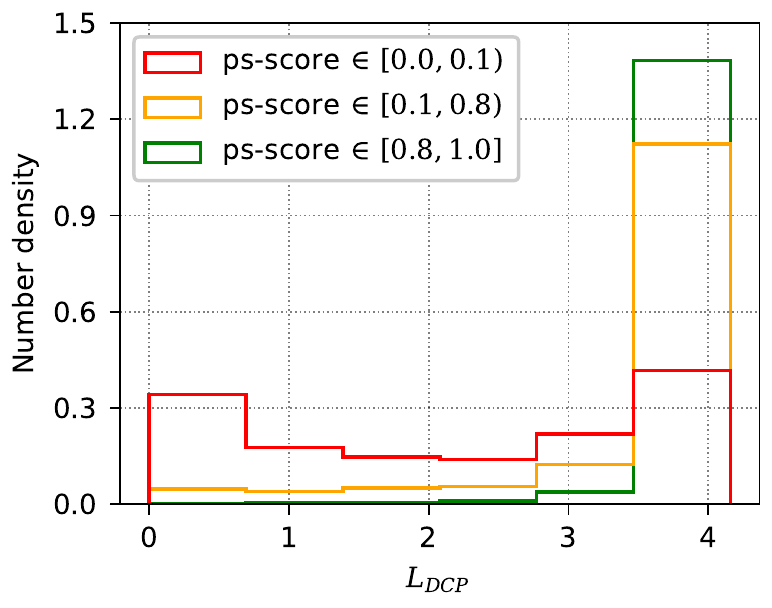}
  \caption{$L_{DCP}$ distribution with respect to different ps-scores, 
  which represent probabilistic inference on being point sources. \label{fig:dcp_loss_wrt_psc}}
\end{figure}

\begin{figure*}
  \includegraphics[width=1.\textwidth]{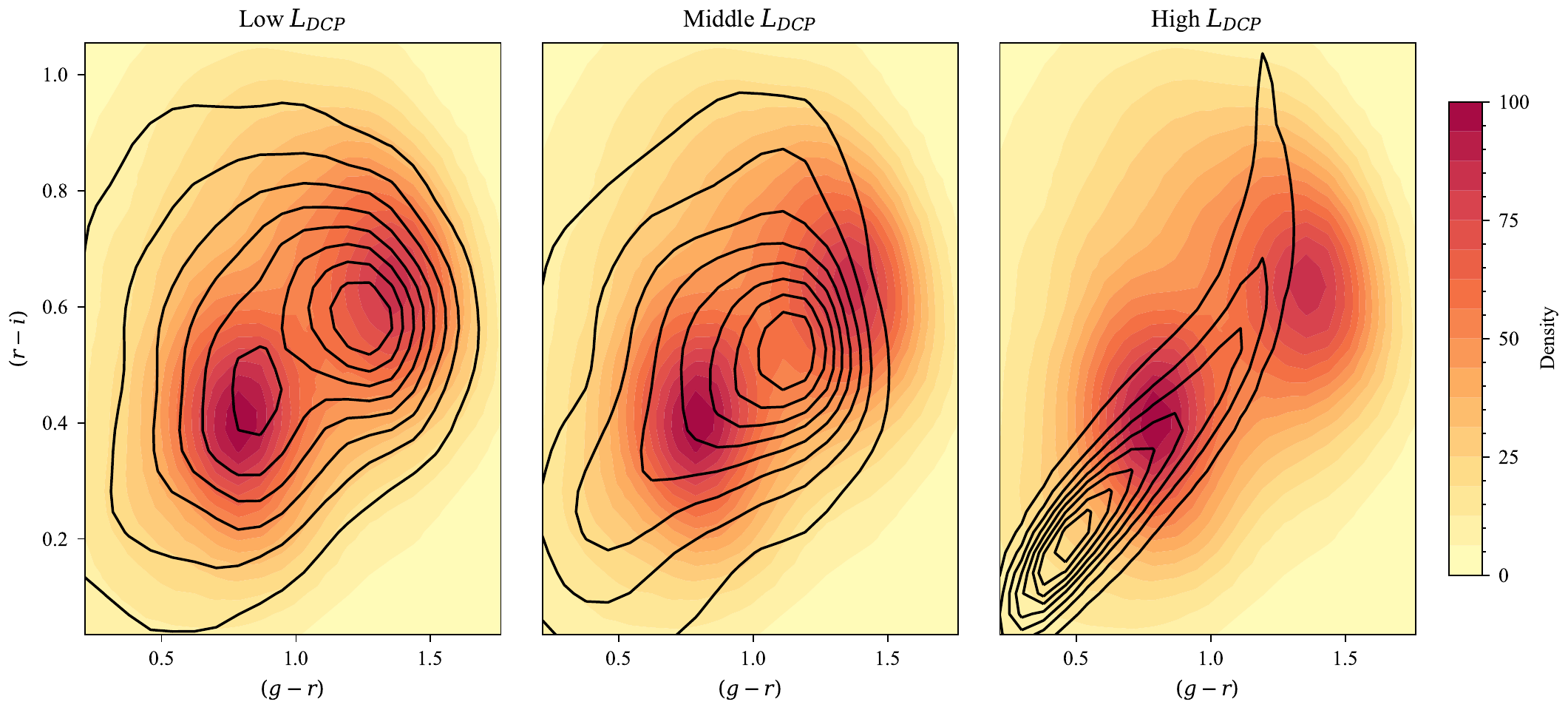}
  \caption{Density contour lines of the UL samples 
    with filled density contour map of the ID samples 
    in the space of input colors $(g-r)$ and $(r-i)$ (PSF measurement). 
    We use the three $L_{DCP}$ ranges to split the UL samples 
    into low, middle, and high $L_{DCP}$ groups 
    as in Figure \ref{fig:red_scatter_cdcp} shows. 
    \label{fig:density_contour}}
\end{figure*}

\begin{figure*}
  \includegraphics[width=1.\textwidth]{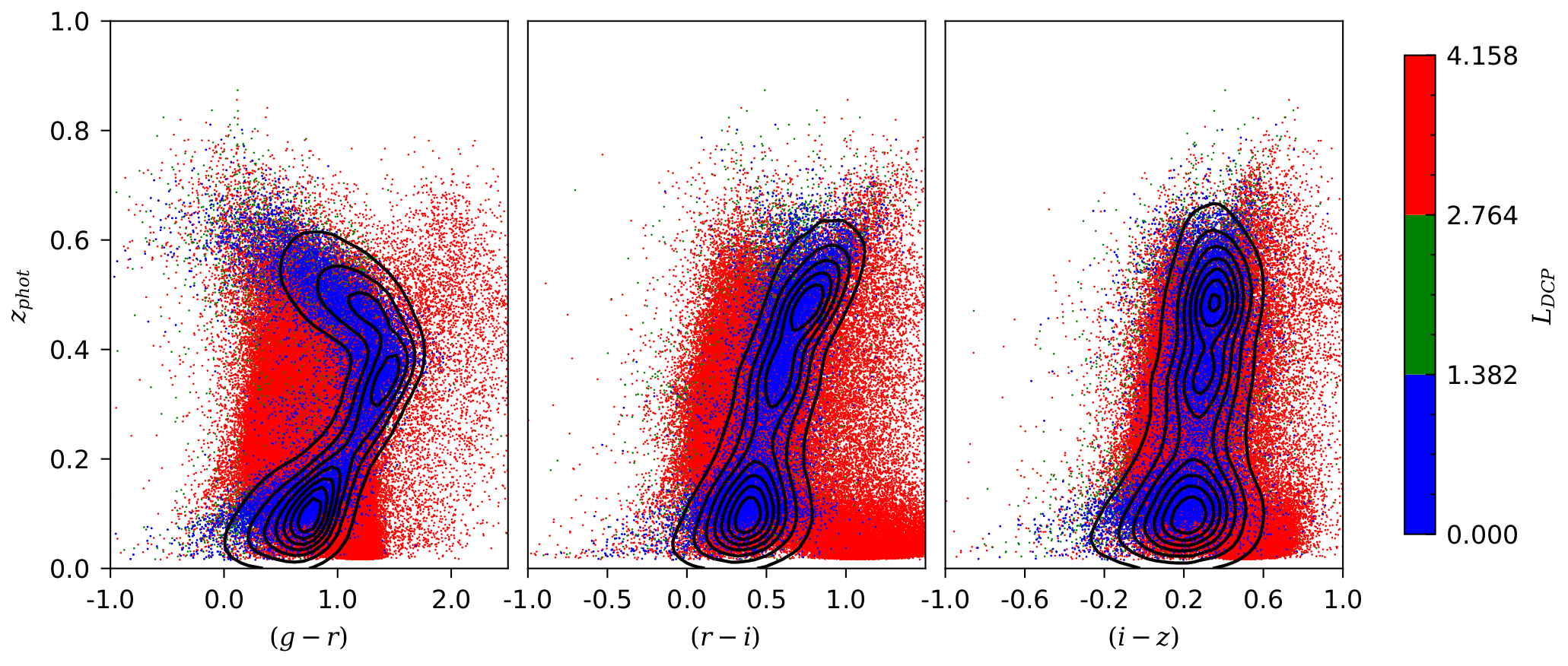}
    \caption{$L_{DCP}$ distributions of the UL samples 
    in the space of input colors (PSF measurement) and estimated photometric redshifts. 
    The black contour lines are drawn with the spectroscopic redshifts and 
    corresponding input colors of the ID samples. 
    The samples used to plot here are randomly under-sampled 
    by a factor of 10 from the entire UL dataset. \label{fig:col_red_scatt_ul}}
\end{figure*}

\begin{figure*}
  \includegraphics[width=1.\textwidth]{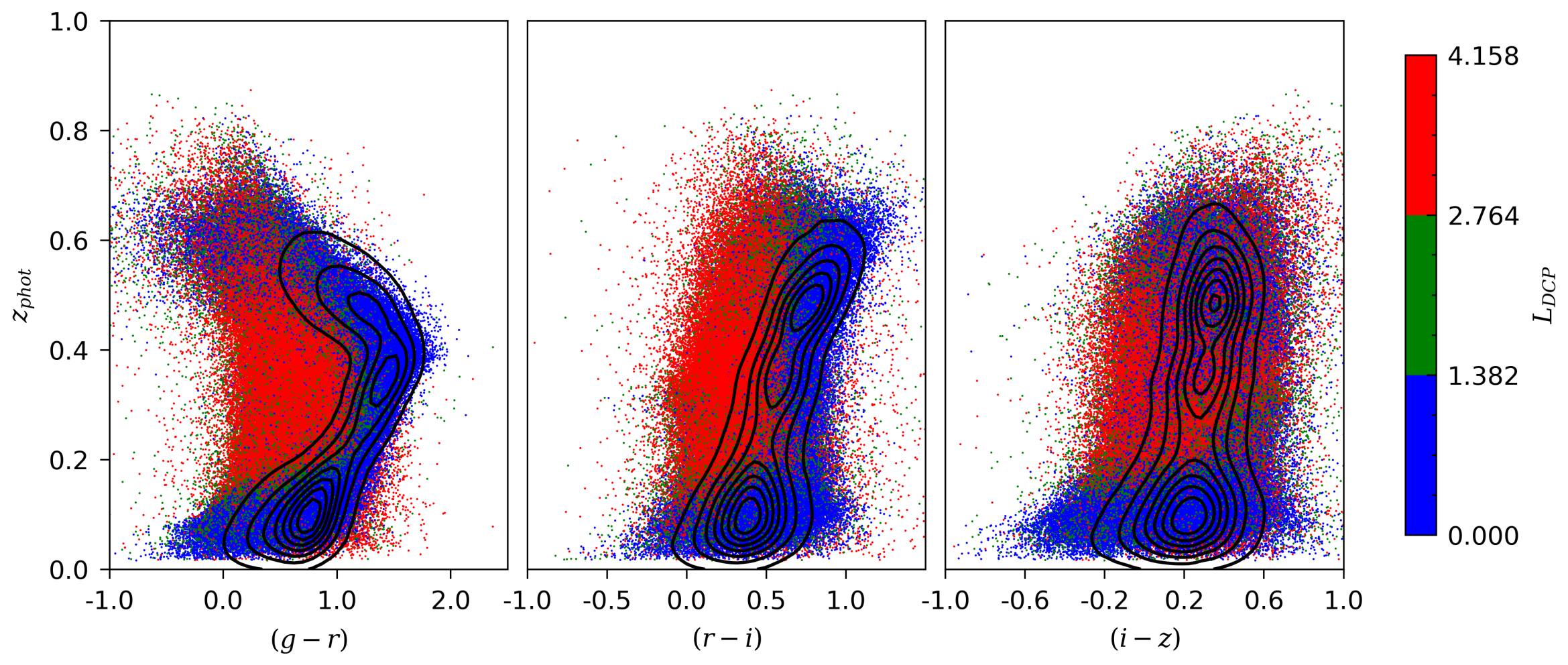}
  \caption{$L_{DCP}$ distributions of the low ps-score ($\le 0.1$) UL samples 
    in the space of input colors (PSF measurement) and estimated 
    photometric redshifts. The black contour lines are drawn with the 
    spectroscopic redshifts and corresponding input colors of the ID 
    samples.
  \label{fig:col_red_scatt_lpsc}}
\end{figure*}

Because the UL data have no labels and spectroscopic redshifts, we 
assess the model performance on the UL samples indirectly. To 
measure the performance of detecting the physically OOD samples 
in the UL dataset, we first extract the point-source score 
(hereinafter, ps-score) from \citet{Tachibana_2018} for the UL samples. 
The ps-score is a machine-learning-based probabilistic inference on 
being point sources for PS1 objects with the range of [0, 1]. 
The samples with low ps-score are more likely to be extended sources, 
i.e., mainly galaxies, and we 
expect our OOD detection algorithm to produce low $L_{DCP}$ 
for samples with low ps-score.

The distributions of $L_{DCP}$ with respect to ps-score are subject 
to the expected pattern. Figure \ref{fig:dcp_loss_wrt_psc} shows 
the $L_{DCP}$ distribution of the samples under-sampled from the UL dataset 
within the RA ranges [0, 10], [100, 110], and [170, 180]. The 
samples within the lowest ps-score interval are mostly distributed 
near the minimum and maximum $L_{DCP}$ values, and the 
distributions of the higher ps-score samples shift toward 
high $L_{DCP}$ ranges. We conjecture that the middle/high $L_{DCP}$ 
samples with low ps-score are highly likely to be 
under-represented galaxies residing in the outskirts or low-density 
parts of the ID distribution (see discussion for Appendix \ref{sec:append}). 
Except for this understandable difference, the two machine-learning 
models reach a consensus on other UL samples which are mainly point-source 
stars and QSOs.

Figure \ref{fig:density_contour} indirectly proves 
$L_{DCP}$ to be a reliable OOD score for a broad range of 
input space for the UL samples. The UL samples with low $L_{DCP}$ have the 
most similar distributions to those of the ID samples because the 
networks assign low $L_{DCP}$ to the UL samples with input 
features comparable with the ID samples. The UL samples 
with middle $L_{DCP}$ begins to display discrepancies from the ID samples 
in input features distributions. 
The samples in the high $L_{DCP}$ group have a completely 
different distribution from the ID samples. In the increasing 
order of the OOD score (i.e., $L_{DCP}$), 
the UL data density distributions in the input color spaces deviate 
more from those of ID samples. As expected from the 
typical colors of stars \citep[e.g.,][]{2019MNRAS.485.4539J}, the 
peak in the color distribution for the high-$L_{DCP}$ UL samples in 
Figure \ref{fig:density_contour} corresponds to common stellar colors.

The training stage-3 model adequately produces photometric 
redshifts for the UL samples. As examined with the ID and LOOD samples, 
we expect the photometric redshift distributions to be similar to those of ID samples 
for the UL samples with low $L_{DCP}$ and random guessing, i.e., uniform 
distribution, for the UL samples with high $L_{DCP}$, respectively. 
As presented in Figure \ref{fig:col_red_scatt_ul}, 
the samples in the low $L_{DCP}$ range mostly reside inside 
the ID sample density contours. 
On the other hand, the photometric redshifts of the OOD-like samples 
randomly spread in the spaces without any strong systematic patterns. 
It certainly shows that the photometric 
redshift distribution for OOD-like samples are closer to a 
uniform distribution than ID-like samples. However, most 
redshifts are in the region $z_{phot} \, < \, 0.8$. It 
arises from the fact that approximately $99.82$\% of the training samples 
for photometric redshifts are found in the region $z_{spec} < 0.8$.

We examine the UL samples with the low ps-score and their $L_{DCP}$ 
in the input feature space. These samples generally correspond to 
galaxies not well represented by the training data. Naturally, the UL samples 
include more faint galaxies than the spectroscopic training samples because 
the procedure of selecting spectroscopic targets is generally biased to bright samples. 
Furthermore, the color and photometric-redshift distributions of 
the UL samples with the low ps-score and high $L_{DCP}$ affirms that 
these UL samples have input features that are not well 
represented by the training samples (see Figure \ref{fig:col_red_scatt_lpsc}).

\section{Discussion and Conclusion \label{sec:discon}}

In this study, we propose a new approach not only to estimate photometric redshifts 
but also to check the reliability of such redshifts through OOD scoring/detection. 
The proposed model successfully estimates the photometric redshifts of galaxies 
using our previous model (presented in Paper I), even with an additional 
network to detect the OOD samples. Our trained model is available with the 
Python code, which is based on PyTorch \citep{NEURIPS2019_9015}, 
on GitHub\footnote{\href{https://github.com/GooLee0123/MBRNN\_OOD}{https://github.com/GooLee0123/MBRNN\_OOD}} 
under an MIT License \citep{lee_and_shin_2021_5611827} for public usage.

The proposed method to detect the OOD data with the photometric 
redshift inference model shows that the new model can measure 
how much OOD data deviate from the training data. 
Most samples with a catastrophic redshift estimation correspond 
to data with high OOD scores (see Figure \ref{fig:red_scatter_cdcp}). 
However, our model estimating the OOD score cannot replace 
models that separate galaxies from stars and QSOs. As shown 
in Figure \ref{fig:dcp_conf_scatt}, numerous stars and QSOs exhibit low 
OOD scores and high confidence values for photometric redshifts in our model 
although they are physically OOD objects. 
These objects have indistinguishable input features compared with galaxies in 
terms of the features used for photometric redshift estimation.

We plan to apply the SED-fitting methods to galaxies 
with high OOD scores or low confidence values. The machine-learning 
inference model has limitations that depend on the training samples. Therefore, 
combining the SED-fitting methods with machine-learning inference 
models can be an effective strategy to achieve high analysis speed and accuracy. 
This would have broad applicability in estimating photometric redshifts for 
numerous galaxies expected for future surveys such as the 
Legacy Survey of Space and Time \citep{2019ApJ...873..111I}.

The current implementation using two networks needs to be 
improved to see the full benefits of ensemble learning. The multiple 
anchor loss parameters presented in Paper I are required for more precise estimation 
of photometric redshifts and evaluation of the OOD scores in a single 
machine-learning framework. The proposed model in this Paper II excludes 
multiple inferences of photometric redshifts in ensemble learning. The ensemble 
model distillation might be a possible technique to accommodate the multiple models 
included in the ensemble learning as a single combined model 
\citep[e.g.,][]{Malinin2020Ensemble} 
in the current framework combining inference of both photometric redshifts 
and OOD score. We plan to develop the implementation of the 
ensemble distillation in the future.

Identifying the influential data among the under-represented galaxies 
in the UL data can play a key role 
in improving the quality of photometric redshifts. Including influential data 
in training can alter the machine-learning model significantly, thereby reducing the fraction 
of incorrect estimation and/or the uncertainty of estimation 
\citep{NEURIPS2019_c61f571d,NEURIPS2020_e6385d39}. 
Because the influence of data relies on the model, 
the future method needs to have a model-dependent algorithm to assess 
the influence of such data quantitatively. The highly influential data 
among the OOD samples requires labeling, i.e., acquiring spectroscopic 
redshifts, to be used as learning samples \citep{2015MasterAJ,NEWMAN201581}.

Understanding the nature of the under-represented galaxies 
will be also crucial in improving the evaluation of the OOD score and 
finding useful samples that need spectroscopic labeling. 
A large fraction of OOD samples might be produced by effects such as source 
blending (including physically merging galaxies) and bias in the 
spectroscopic target selection. When selecting highly 
influential OOD samples among the UL data, our new algorithm for the OOD score 
may need a step of filtering out the photometrically unreliable samples such as 
blended galaxies. The under-represented galaxies might include galaxies 
with less common physical properties; e.g., extreme dust obstruction, 
unusual stellar populations, short post-starburst phase, unusually intensively 
star-forming galaxies, merging galaxies at various evolutionary stages, 
and galaxies with weak QSO emission.

Semi-supervised learning methods 
might be an alternative approach to the supervised learning 
such as our proposed approach. 
Although semi-supervised learning has its own limitations such as 
smoothness assumption \citep{10.5555/1208768}, semi-supervised learning models 
intrinsically do not have the OOD problem, except for 
physically OOD samples. However, the 
efficacy and reliability of the semi-supervised learning is a new challenge 
compared with supervised learning methods. Measuring the influence of 
data on the training and implementation of the algorithm is crucial 
because the performance of semi-supervised learning 
methods is strongly subjective to the labeled samples and the weights of 
UL samples.

\begin{acknowledgments}

We thank David Parkinson for his careful reading and thoughtful comments. We are 
also grateful to Kyungmin Kim and Hyung Mok Lee for insightful discussions. 
We also thank the anonymous referee for helpful comments 
with valuable suggestions. 
This research has made use of the VizieR catalog access tool, CDS,
Strasbourg, France (DOI : 10.26093/cds/vizier). The original description 
of the VizieR service was published in 2000, A\&AS 143, 23. 
The Pan-STARRS1 Surveys (PS1) and the PS1 public science archive have been made 
possible through contributions by the Institute for Astronomy, the University 
of Hawaii, the Pan-STARRS Project Office, the Max-Planck Society and its 
participating institutes, the Max Planck Institute for Astronomy, Heidelberg 
and the Max Planck Institute for Extraterrestrial Physics, Garching, The Johns 
Hopkins University, Durham University, the University of Edinburgh, the Queen's 
University Belfast, the Harvard-Smithsonian Center for Astrophysics, the Las 
Cumbres Observatory Global Telescope Network Incorporated, the National Central 
University of Taiwan, the Space Telescope Science Institute, the National 
Aeronautics and Space Administration under Grant No. NNX08AR22G issued through 
the Planetary Science Division of the NASA Science Mission Directorate, 
the National Science Foundation Grant No. AST-1238877, the University of Maryland, 
Eotvos Lorand University (ELTE), the Los Alamos National Laboratory, and 
the Gordon and Betty Moore Foundation.

\end{acknowledgments}

\appendix

\setcounter{table}{0}
\renewcommand{\thetable}{\Alph{section}.\arabic{table}}
\setcounter{figure}{0}
\renewcommand{\thefigure}{\Alph{section}.\arabic{figure}}

\section{Example of under-represented galaxies in training data and $L_{DCP}$ \label{sec:append}}

We demonstrate a typical case of selection effects based on brightness 
in spectroscopic training data as an example of considering 
OOD data. 
Since the brightness of objects is a main factor affecting target selection 
in spectroscopy, we examine an extreme case when training samples of 
spectroscopic redshifts consist of only bright objects, 
and data for inference include only faint objects. This case 
undoubtedly demonstrates our model to detect OOD data 
works well.

\begin{figure*}
\centering
\includegraphics[width=0.49\textwidth]{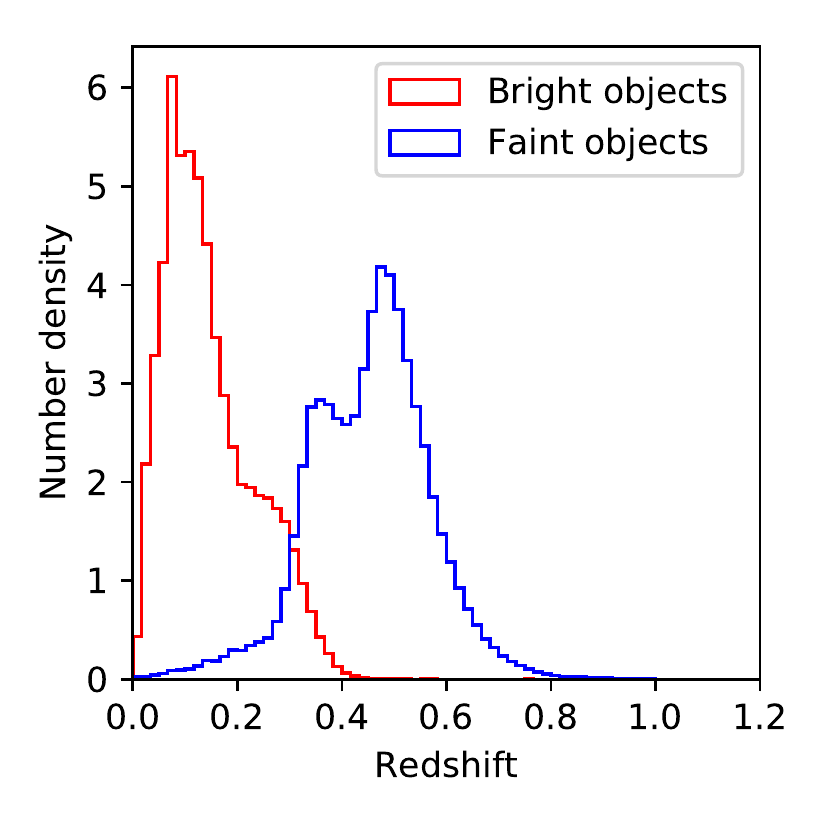}
\includegraphics[width=0.49\textwidth]{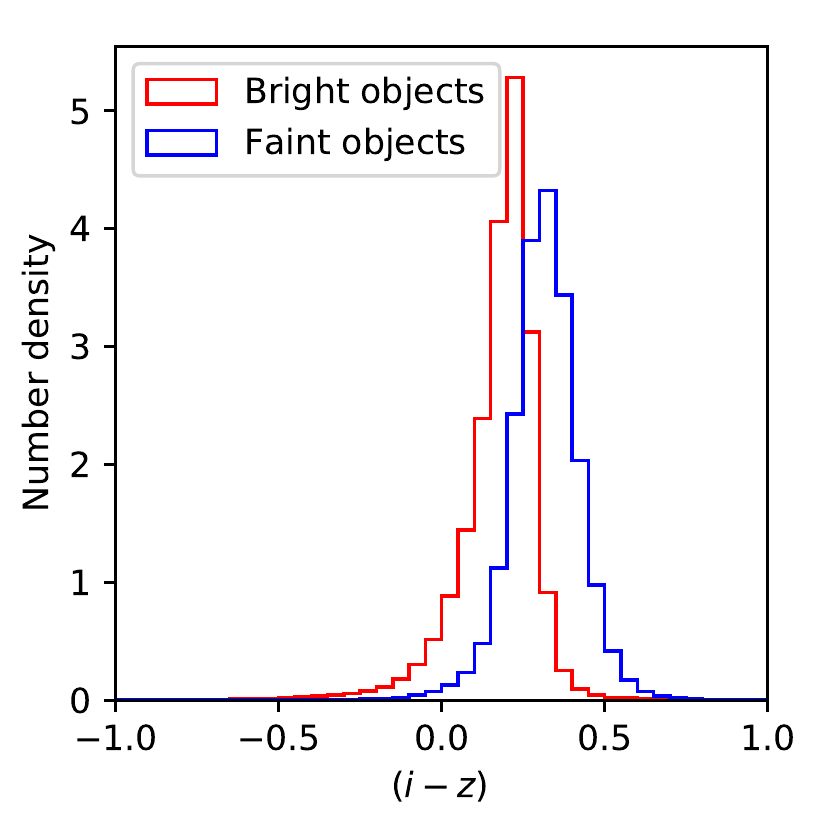}
\caption{Distribution of spectroscopic redshifts ({\it left}) 
and $(i - z)$ in Kron measurement ({\it right}) for bright and faint objects in 
the SDSS training data. Bright objects have $r$-band Kron magnitude $\le$ 19. 
\label{fig:appendix_redshift_color}}
\end{figure*}

Adopting the neural network model in the same way 
as described in Section \ref{sec:method}, 
we train the model with the bright SDSS spectroscopic objects, which 
have $r$-band Kron magnitude less than or equal to 19, as supervised training 
samples. The properties of the 634,928 bright SDSS objects are different from 
those of the 659,114 faint objects as expected in the SDSS data 
(see Figure \ref{fig:appendix_redshift_color}). The bright objects generally 
have lower redshifts and bluer color than the faint objects have. Therefore, 
a large fraction of the faint objects should be OOD 
objects with respect to the trained model as under-represented galaxies in 
the training data.

\begin{figure*}
\centering
\includegraphics[width=1.\textwidth]{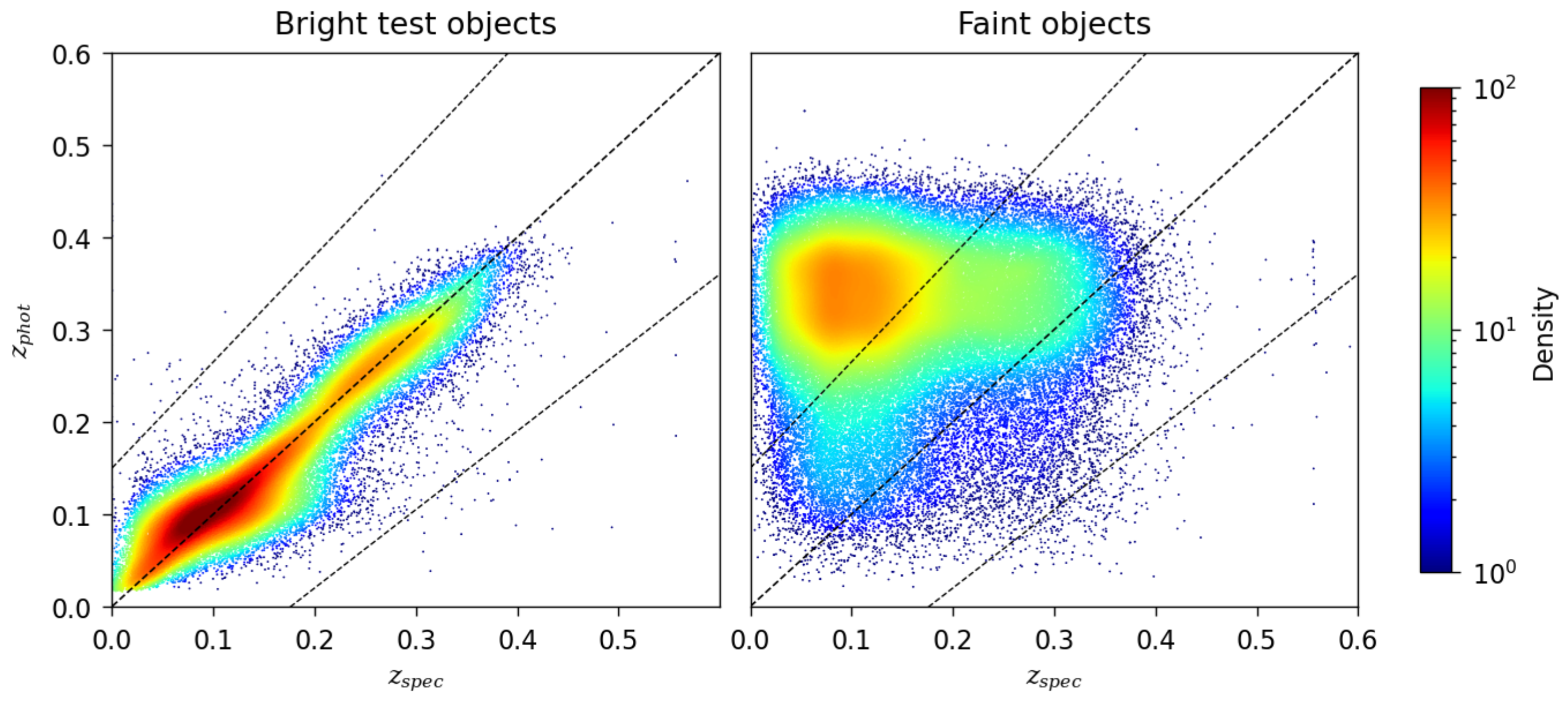}
\caption{Comparison between spectroscopic and photometric redshifts 
for the bright test objects (i.e., ID data; {\it left}) and the faint objects 
(i.e., OOD candidates; {\it right}). The scatters are color-coded according to 
number density. The dashed lines correspond to slope-one lines and catastrophic 
error boundaries, respectively.
\label{fig:appendix_specz_photoz}}
\end{figure*}

\begin{figure*}
\centering
\includegraphics[width=0.5\textwidth]{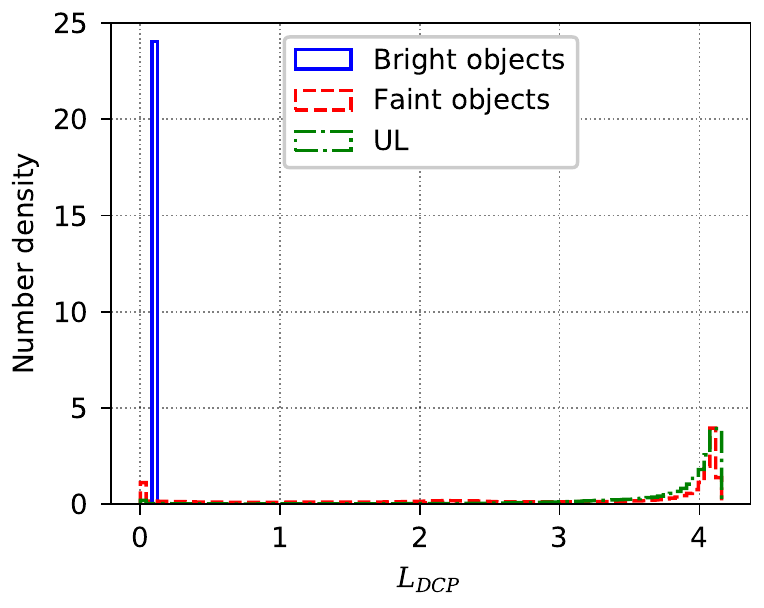}
  \caption{$L_{DCP}$ distribution of the bright test objects, faint objects, 
  and UL samples. The bright test objects are basically ID data, but 
  the majority of faint objects and UL samples have 
  high $L_{DCP}$ corresponding to the OOD candidates.
\label{fig:appendix_ldcp}}
\end{figure*}

\begin{figure*}
\centering
\includegraphics[width=1.\textwidth]{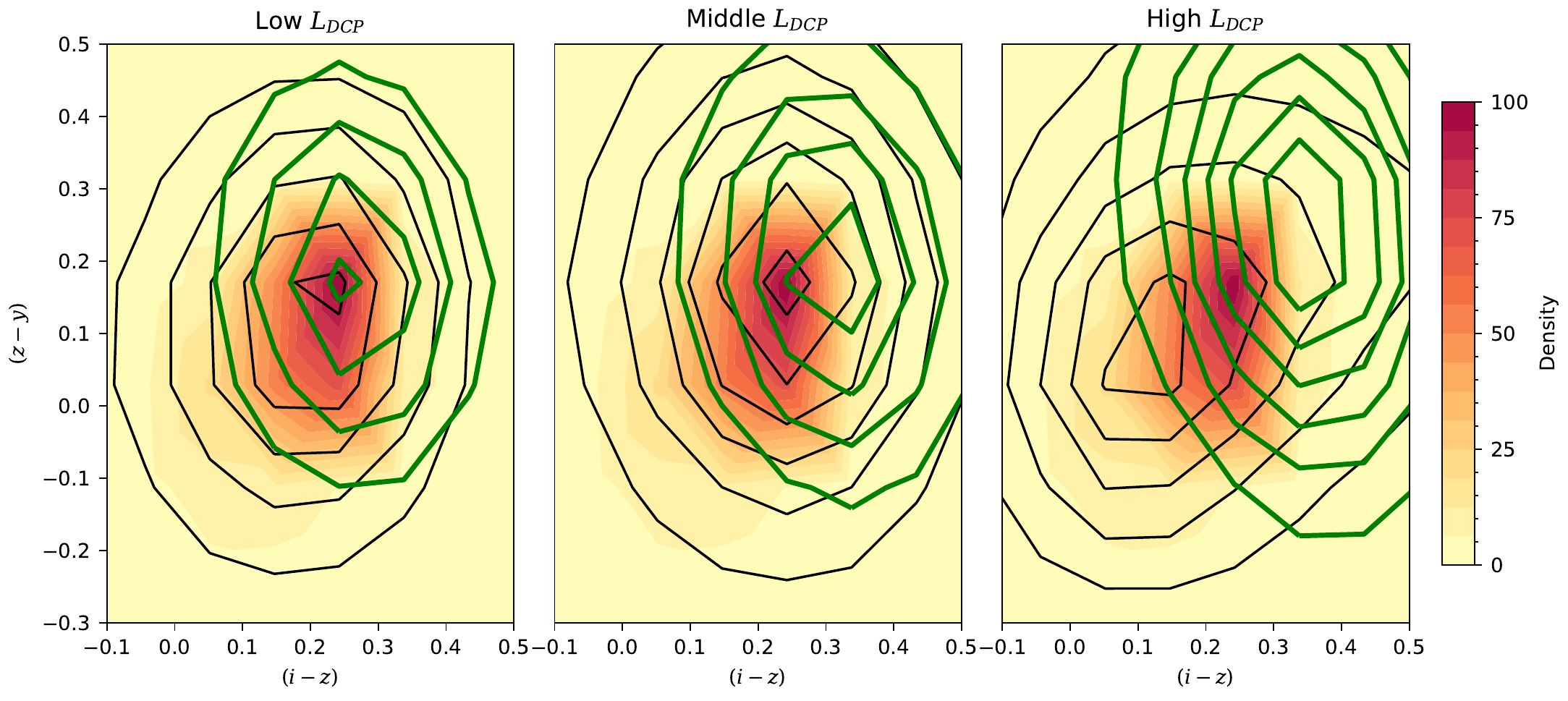}
  \caption{Distribution of $(i - z)$ and $(z - y)$ colors (Kron measurement) 
  with respect to the 
  three different ranges of $L_{DCP}$ values, which are the same as Figure \ref{fig:red_scatter_cdcp}, 
  for the faint objects and UL samples. 
  The filled contours correspond to the bright test samples (i.e., the 
  ID samples). The black and green contour lines represent 
  the UL and faint samples, respectively.
\label{fig:appendix_color}}
\end{figure*}

The performance in estimating photometric redshifts is much worse for the 
faint objects than for the bright test samples. As shown in 
Figure \ref{fig:appendix_specz_photoz}, the trained model cannot 
estimate correct photometric redshifts for the majority of the faint objects 
because they are under-represented objects, i.e., OOD objects. 
The point-estimation metrics also show that the model is not able to infer 
photometric redshifts for the faint objects. For example, $R_{cat}$ values 
are 0.0012 and 0.22 for the bright test samples and the faint objects, respectively. 
The bias value for the bright test samples is about 0.004 while that value 
for the faint objects is about 0.09.

Our method to evaluate $L_{DCP}$ works well in this demonstration. 
Figure \ref{fig:appendix_ldcp} shows that our method 
successfully gives the faint objects high $L_{DCP}$. The training 
data composed of the bright objects do not cover a substantial range of 
redshifts and input features for the faint test objects as presented 
in Figure \ref{fig:appendix_redshift_color}. Most of the faint objects have 
high $L_{DCP}$ as we expect. Because most UL samples are stars in our Galaxy 
or faint objects, they also show high $L_{DCP}$ values 
in Figure \ref{fig:appendix_ldcp}.

Colors as well as their high uncertainties of the faint SDSS objects, which 
have the high $L_{DCP}$ values, display distinct deviation 
from those of the bright spectroscopic samples (see Figure \ref{fig:appendix_color}). 
Because we use $r$-band magnitude 
to divide the SDSS spectroscopic samples into the bright and faint sets, 
redder objects are more likely to be OOD objects 
as faint objects. As presented in Figure \ref{fig:appendix_color}, 
the faint spectroscopic galaxies showing high $L_{DCP}$ have 
significantly redder $(i - z)$ and $(z - y)$ colors 
than the training samples used to derive the model.

The colors of the UL samples with high $L_{DCP}$ are clearly different 
from those of the faint galaxies with high $L_{DCP}$. In Figure 
\ref{fig:appendix_color}, the peak of the color distribution for the UL 
samples with high $L_{DCP}$ is much bluer than those of both the 
bright and faint SDSS galaxies 
with high $L_{DCP}$. Considering the typical distribution of 
colors for stars, QSOs, and galaxies in the SDSS filter bands 
\citep[e.g.][]{2000AJ....120.2615F,2012MNRAS.427.2376H,2019MNRAS.485.4539J}, 
the peak in the color distribution for the UL samples corresponds 
to the expected colors of stars and low-redshift QSOs.

\bibliography{astro,ml}{}
\bibliographystyle{aasjournal}

\end{document}